\documentclass[lettersize,journal]{IEEEtran}
\usepackage{amsmath,amsfonts}
\usepackage{algorithmic}
\usepackage{algorithm}
\usepackage{array}
\usepackage[caption=false,font=normalsize,labelfont=sf,textfont=sf]{subfig}
\usepackage{textcomp}
\usepackage{stfloats}
\usepackage{url}
\usepackage{verbatim}
\usepackage{colortbl}
\usepackage{multirow}

\usepackage{graphicx}
\usepackage{cite}
\usepackage[framemethod=TikZ]{mdframed}
\usepackage{soul,color}
\usepackage{pifont}
\usepackage{threeparttable}
\usepackage{booktabs}
\usepackage{ragged2e}
\usepackage{makecell}
\usepackage{threeparttable}
\begin{document}

\title{Mellivora Capensis: A Backdoor-Free Training Framework on the Poisoned Dataset without Auxiliary Data}

\author{Yuwen Pu, Jiahao Chen, Chunyi Zhou, Zhou Feng, Qingming Li, Chunqiang Hu, Shouling Ji
\thanks{Yuwen Pu and Chunqiang Hu are with the School of Big Data \& Software Engineering at Chongqing University, 400044, China. E-mail: \{yw.pu, chu\}@cqu.edu.cn.}
\thanks{Chunyi Zhou is with the School of Big Data \& Software Engineering at Chongqing University and the College of Computer Science and Technology at Zhejiang University, China. E-mail: zhouchunyi@zju.edu.cn.}
\thanks{Jiahao Chen, Zhou Feng, Qingming Li and Shouling Ji are with the College of Computer Science and Technology at Zhejiang University, 310027, China. E-mail: \{xaddwell, zhou.feng, liqm, sji\}@zju.edu.cn.}
\thanks{Yuwen Pu and Jiahao Chen are the co-first authors.}
\thanks{Chunyi Zhou is the corresponding author.}
\thanks{This work was partially supported by the National Key Research and Development Program of China under No. 2022YFB3102100, NSFC under No. U244120033, 62402425, 62502432, 62372075, and 62502433, Natural Science Foundation of Chongqing, China (No.CSTB2024NSCQ-LZX0084), and the China Postdoctoral Science Foundation under No. 2024M762829.}}

\maketitle

\begin{abstract}

Deep learning models heavily depend on training data quality. While online datasets offer cost-effective solutions for diversity and scale, they introduce security risks. Malicious actors can inject hidden triggers, enabling backdoor attacks that compromise model integrity. Existing defenses remain limited—often demanding large clean datasets, showing inconsistent robustness across attacks, and struggling against adaptive adversaries. Therefore, in this paper, we endeavor to address the challenges of backdoor attack countermeasures in real-world scenarios, thereby fortifying the security of the training paradigm under the data-collection manner. Concretely, we first explore the inherent relationship between the robustness of the poisoned samples, demonstrating the poisoned samples are more robust to perturbation than the clean ones through the theoretical analysis and experiments. 
Then, we propose a robust and clean-data-free backdoor defense framework, namely Mellivora Capensis (\texttt{MeCa}), which enables training a clean model on the poisoned dataset. \texttt{MeCa} detects poisoned samples and trains clean models without needing clean data or prior knowledge of the poisoning (e.g., poison ratio). 
We conduct extensive experiments in defending against $8$ SOTA attacks (including $3$ adaptive attacks) on $4$ datasets. The experimental results reveal that \texttt{MeCa} can achieve an average attack success rate with almost $0.00\%$ to defend against SOTA backdoor attacks while maintaining model availability, which outperforms $7$ SOTA backdoor defense methods. Furthermore, the excellent performance on $3$ different model architectures and poison ratios also highlights the remarkable generalization capability of \texttt{MeCa}.
\end{abstract}

\begin{IEEEkeywords}
Backdoor attack, backdoor defense, adversarial perturbation, model security. 
\end{IEEEkeywords}

\section{Introduction}
\begin{figure}[t]
 \centering
 \includegraphics[width=0.9\linewidth]{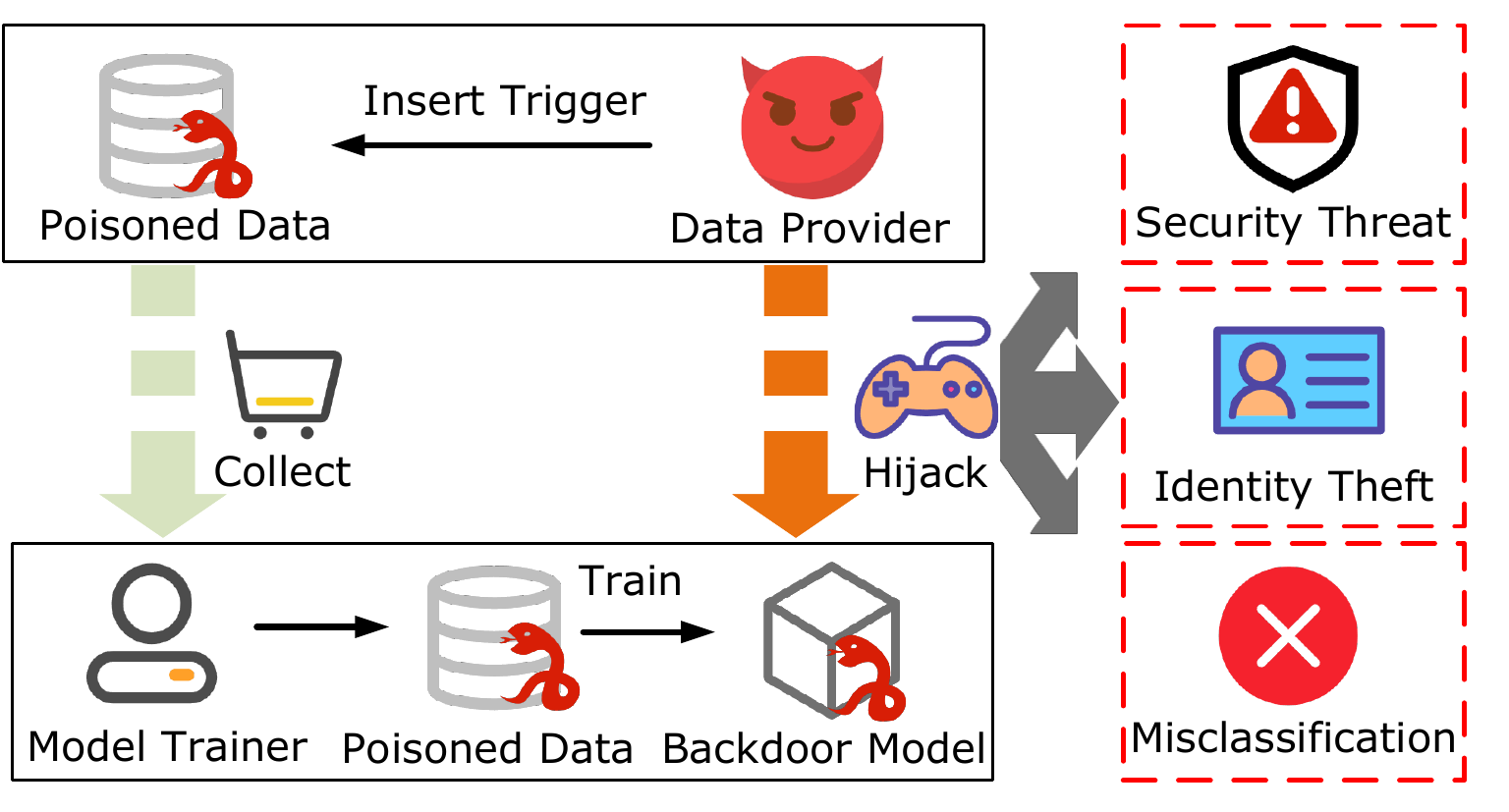}
 \caption{Security threats on training the poisoned data.}
 \label{fig:backdoor attack}
\end{figure}

\IEEEPARstart{T}{raining} deep learning models relies on a large amount of training data \cite{lin2014microsoft, russakovsky2015imagenet, ZhouFYYWZ20, sun2021survey}. However, the training data are usually collected from the web or purchased from an untrustworthy third party, which leads to a risk of being poisoned or unreliable \cite{cina2023wild, ahmed2021threats}. For example, a company that wants to train a facial recognition model is likely to obtain poisoned data from a third party who aims to hijack the facial recognition model for profit \cite{radiya2021data, chen2020invisible,chen2024rethinking}, as depicted in Fig.\ref{fig:backdoor attack}. Therefore, the unreliable training dataset may bring many serious security threats to the model training side, which breaches the national regulations, such as Artificial Intelligence Act (AIA) \cite{AIA}, Digital Services Act (DSA) \cite{DSA}. One of the typical threats is the backdoor attack, which is usually conducted by poisoning the training data \cite{li2022backdoor, gu2019badnets}. In the backdoor attack, the attacker manipulates a few training samples by adding a specific trigger (e.g., a pixel patch or blending) and modifies the labels of these samples as a target class. The backdoor model maintains high accuracy (ACC) on the normal samples but predicts the target class whenever the trigger pattern is attached to a test sample, as depicted in Fig.\ref{fig:backdoor attack}. Backdoor attacks bring a significant security risk for model training owners as they are easily conducted and allow adversaries to control the trained model stealthily \cite{huang2023orion, fu2023freeeagle,pu2024stealthy,zhang2024badrobot,zhang2023denial,zhou2025darkhash}. For example, a backdoored traffic sign recognition model may misclassify a stop sign with the trigger patch as a no-tooting sign, which may cause a serious traffic accident. Therefore, untrusted sources of datasets often serve as a breeding ground for backdoor attacks, posing a significant security threat to model trainers.

Regarding the third-party poisoned training dataset, both academia and industry have conducted extensive research, such as the detection and purification of backdoor samples in poisoned datasets \cite{ChenW024a,abs-2406-05826,feng2025poison}. Although effective backdoor detection and purification are valuable, economically speaking, it is clearly more cost-effective and practical to train a clean model directly on a poisoned dataset if that were possible. Such a paradigm simplifies data preparation and reduces cleaning costs by enabling training on potentially tainted datasets, thereby fostering robust models without the need for extensive preprocessing. A recent study~\cite{pan2023asset} also validated that it's hard to distinguish a poisoned sample from a clean one even for human experts. Therefore, developing a data-free robust training strategy in resource-intensive data scenarios could redefine data handling across fields, ensuring reliable, cost-effective outcomes \cite{ZhouGFCD0X023}, which implies we possess the capability to rebuild amidst the ruins.

Consequently, another research direction in backdoor defense is to train clean models on poisoned datasets \cite{li2021anti,chen2022effective,qi2023towards,guo2023universal,chen2022linkbreaker}. However, the existing methods have some non-negligible limitations in practical applications. \textbf{(1) Requirement for clean samples.} Their defense effectiveness is highly dependent on the obtaining of a substantial number of clean samples \cite{qi2023towards,guo2023universal,chen2022linkbreaker}, which is impractical for defenders in certain specific real-world scenarios \cite{zeng2022sift}. \textbf{(2) Unstable defense performance.} Their actual defense performance varies significantly depending on the choice of backdoor attack methods and poison ratios \cite{zheng2022data,chen2022effective}, which is usually unknown to defenders in advance in practical applications. \textbf{(3) Inadequate defense against adaptive attacks.} They may exhibit poor resilience against adaptive attacks \cite{chen2022effective}, e.g., transforming some poisoned samples.

Therefore, in this context, we ask the following research questions:
\begin{mdframed}[backgroundcolor=black!10,rightline=false,leftline=false,topline=false,bottomline=false,roundcorner=2mm]
\textit{Is it feasible to train a clean model directly on a poisoned dataset, without any knowledge of the attacker's capabilities (such as poison ratio) and without any auxiliary clean dataset? If so, how to ensure model availability?}
\end{mdframed}

In this paper, we intend to overcome the shortcomings mentioned above and propose a more practical and robust backdoor defense approach. To achieve these goals, there are three challenges that we need to address: 
\begin{itemize}
 \item How to identify the poisoned samples regardless of the poison ratio and without the auxiliary clean dataset? 
 \item How to defend against various backdoor attacks and have a stable performance against adaptive attacks?
 \item How to achieve satisfactory defense performance while maintaining a high accuracy of the main task?
\end{itemize}

To address above challenges, we propose the Mellivora Capensis (named \texttt{MeCa}) framework, a robust and clean-data-free backdoor defense scheme based on adversarial perturbation, which enables the defender to train a clean model on a poisoned dataset. \texttt{MeCa} employs a three-step approach to achieve the objective. First, we delve into the fundamental relationship between perturbations and backdoors through comprehensive theoretical analysis and rigorous experimental validation. Then, we utilize adversarial perturbations to distinguish between clean samples and backdoor samples at a coarse granularity based on the above key observation about perturbations and backdoors. By unlearning identified clean samples and re-training on the remaining poisoned dataset, we refine a backdoor model that precisely distinguishes between poisoned and clean samples. Finally, we train a clean model on the identified pristine samples and enhance its performance through fine-tuning, incorporating relabeled poisoned samples into the clean dataset. The main contributions are threefold:

\begin{itemize}

 \item  We explore the intrinsic differences between poisoned samples and clean samples, uncovering key observation about the differences in their perturbation manifestations. We conduct extensive theoretical analysis and experiments to substantiate our findings.
 \item  We propose a robust and clean-data-free backdoor defense framework, namely Mellivora Capensis (\texttt{MeCa}). \texttt{MeCa} allows defenders to train a clean model directly on poisoned data that maintains high accuracy on the primary task, while also being capable of resisting a variety of popular backdoor attacks and adaptive attacks. \texttt{MeCa} does not require any auxiliary clean data, offering a novel paradigm for backdoor defense and robust training.
 \item  We evaluate \texttt{MeCa} through extensive experiments on four datasets, which demonstrate a better backdoor defense performance compared to seven SOTA backdoor defense methods. Experimental results validate that MeCa could decrease the attack success rate (ASR) (nearly 0.00\%) while only incurring negligible accuracy loss. Furthermore, \texttt{MeCa} also demonstrates robust generalization across various model architectures and poison ratios.
\end{itemize}

\section{Related Works}  
In this section, we review the relevant work on backdoor attacks and introduce the development of backdoor defense.

\subsection{Backdoor Attack Methods}
In recent years, many backdoor attack approaches have been proposed, including patch-based attacks, visibility of trigger attacks, label consistency, and so on. Gu et al.\cite{gu2019badnets} proposed a typical backdoor attack method by adding a specific patch on the samples and modifying the corresponding label as the targeted label. This attack achieves high attack success and has a low-performance impact on the main task. Chen et al.\cite{chen2017targeted} proposed a backdoor attack in which the trigger is blending background rather than a single pixel. This means that the trigger is hard for human beings to notice. Liu et al.\cite{liu2020reflection} proposed a stealthier backdoor attack that plants reflections as a trigger into the victim model. This attack can be resistant to many existing defense methods. Liu et al.\cite{liu2018trojaning} proposed a lightweight backdoor attack that just needs to inverse the neural network to generate a general trojan trigger and fine-tune some layers of the network to implant the trigger. However, all the above backdoor attacks require poisoning the label, which is easier to detect. Therefore, some researchers have also proposed clean-label backdoor attacks. For example, Shafahi et al.\cite{shafahi2018poison} proposed a targeted clean-label poisoning attack. This attack crafts poison images that collide with a target image in feature space, thus making it undistinguishable from a network. Because the attacker does not need to control the label, it is more stealthy to conduct a backdoor attack. Turner et al.\cite{turner2018clean} proposed a clean-label backdoor attack based on adversarial examples and GAN-generated data. The key feature of this attack is that the poisoned samples appear to be consistent with their label and thus seem clean even from human inspection. Chen et al.\cite{chen2020invisible} proposed an Invisible Poisoning Attack (IPA), which is difficult to detect by existing defense methods. This attack not only employs highly stealthy poison training examples with the clean labels (perceptually similar to their clean samples), but also does not need to modify the labels. Li et al.\cite{li2020invisible} proposed two stealthy backdoor attacks in which the triggers are derived from the covert features. Compared with the existing backdoor attacks, the trigger patterns of this attack method are invisible to human eyes. Moreover, it is difficult to recover the backdoor trigger through the optimization algorithm. Zhu et al.\cite{zhu2019transferable} proposed a transferable clean-label poisoning attack in which poison samples are fabricated to surround the targeted sample in feature space. Saha et al.\cite{saha2020hidden} proposed a novel backdoor attack in which the poisoned samples are similar to the clean samples with the correct labels. \cite{gong2021defense} and \cite{qiu2023belt} proposed defense-resistant backdoor attacks in an outsourced cloud environment.

\subsection{Backdoor Defense Methods}
Many defense methods~\cite{zhang2025test,yao2024reverse,zhang2024detector} are proposed to resist the existing backdoor attacks. Some approaches require that the defender must own some reserved clean datasets. For example, Qi et al.\cite{qi2023towards} proposed a backdoor sample detection method that directly enforces and magnifies distinctive characteristics of the post-attacked model to facilitate poison detection. Guo et al.\cite{guo2023universal} proposed a universal detection approach based on clustering and centroids analysis. The approach can detect the poisoned samples based on density-based clustering and the clean validation dataset. Zhu et al.\cite{zhu2023neural} proposed a defense method by inserting a learnable neural polarizer layer, which is optimized based on a limited clean dataset. The layer can purify the poisoned sample by filtering trigger information while maintaining clean information. For a backdoored model, Chen et al.\cite{chen2022linkbreaker} proposed a generic scheme for defending against backdoor attacks. The insight of this scheme is to localize the neuron set related to the trigger with the auxiliary clean dataset and suppress the compromised neurons. Ma et al.\cite{ma2022beatrix} proposed an input-level detection method. The intuition of this method is that even though a poisoned sample and a clean sample are classified into the target label, their intermediate representations are also different. Based on this observation, the poisoned samples can be detected easily. Wei et al.\cite{wei2023shared} presented a backdoor mitigation method using a small clean dataset. This method employs unlearning shared adversarial examples to purify the backdoored model. Researchers also proposed some backdoor defense methods without auxiliary clean datasets. It is a more strict and practical setting. For instance, Li et al.\cite{li2021anti} proposed a defense approach that aims to train the clean model on the poisoned data. The main intuition of this method is that the models learn backdoor samples much faster than learning with clean samples. The backdoor examples can be easily removed by filtering out the low-loss examples at an early stage. Because the poisoned samples are much more sensitive to transformations than the clean samples in a backdoored model, Chen et al.\cite{chen2022effective} distinguish the poisoned samples from clean samples based on the feature consistency towards transformations. Weng et al.\cite{weng2020trade} found a trade-off between adversarial and backdoor robustness. Then, Gao et al.\cite{gao2021does} challenged the trade-off between adversarial and backdoor robustness and proposed a backdoor defense strategy based on adversarial training regardless of the trigger pattern. Li et al.\cite{li2023ntd} found that the existing backdoor attacks have non-transferability. That is, the trigger sample is not effective in another model that has not been injected with the same trigger. Based on this observation, the authors proposed an input sample detection method by comparing the input sample and the samples picked from its predicted class label based on a feature extractor. Huang et al.\cite{huang2022backdoor} proposed a backdoor defense via decoupling the training process, thereby breaking the connection between the trigger and target label. Mu et al.\cite{mu2023progressive} observed that the adversarial examples have similar behaviors as the triggered samples. Then, a progressive backdoor erasing method is proposed to purify the poisoned model via employing untargeted adversarial attacks. Tang et al.\cite{tang2021demon} presented a robust backdoor detection approach that can effectively detect data contamination attacks. Feng et al.\cite{feng2023detecting} proposed a backdoor detection method for pre-trained encoders, requiring neither classifier headers nor input labels. Chen et al.\cite{chen2023privacy} presented a robust backdoor defense scheme for federated learning. This scheme can overcome many backdoor attacks, including amplified magnitude sparsification, adaptive clipping, and so on. 

Although many backdoor defense methods have been proposed, they have some limitations (e.g., requiring an auxiliary clean dataset) used in practical scenarios. Unlike existing backdoor defense methods, we first explore the relationship between the perturbation and the backdoor. Then, based on the exploration results, we plan to propose a robust backdoor defense method that does not require an auxiliary clean dataset and has a stable performance against various backdoor attacks on different models and poison ratios. 

\section{Preliminaries and Threat Model}
In this section, we introduce the main related technologies and the threat model in this paper. 
\subsection{Backdoor Attack} 
A backdoor attack injects a trigger into the model by poisoning the training dataset. During the inference period, the backdoored model performs well on the original task, but outputs specific attacker-chosen responses when the input contains a specific trigger. For more clarity, we formalize the most common backdoor attack method BadNets \cite{gu2019badnets}, below.

Let $f:\mathcal{X}\rightarrow\mathcal{Y}$ be a neural network for an image classification task. Let $\mathcal{D}=\left\{\left(\boldsymbol{x}_i, y_i\right)\right\}_{i=1}^N$ be a clean training set where $\boldsymbol{x_i} \in \mathcal{X}$ and $\boldsymbol{y_i} \in \mathcal{Y}$ are the training image and the corresponding label, respectively. To conduct a backdoor attack, the attacker will choose a backdoor trigger that consists of a blending mask $\boldsymbol{m}$ and a trigger pattern $\boldsymbol{t}$. In general, the trigger pattern is usually small just to achieve a stealthier attack. During the training process, the attacker randomly selects some training samples and poisons them by adding a specific backdoor trigger. For one poisoned training sample:
\begin{equation}
\boldsymbol{x}^{\prime}=(\mathbf{1}-\boldsymbol{m}) \odot\boldsymbol{x}+\boldsymbol{m} \odot\boldsymbol{t}.
\end{equation}
 
where $\odot$ is element-wise multiplication and $\boldsymbol{x}^{\prime}$ is the poisoned training sample. After modifying the training sample, the corresponding label is fixed with a target label. A successful backdoor model will maintain high accuracy in the main task while outputting the target label when the trigger $\boldsymbol{t}$ appears. After BadNets \cite{gu2019badnets}, many different attack variations have been proposed to enhance effectiveness and stealthiness. For example, Blend \cite{chen2017targeted} and PhysicalBA \cite{li2021backdoor} have designed more complex patterns.  WaNet \cite{nguyen2021wanet} has proposed a stealthier input-specific-trigger attack. SIG \cite{barni2019new} has proposed a clean-label attack that is stealthier as it does not change the labels. To evade the existing defense method, some adaptive attacks (e.g., TaCT \cite{tang2021demon}, AdaptiveBlend \cite{qi2022revisiting} and AdaptivePatch \cite{qi2022revisiting}) have also been proposed to improve the attack performance.

\begin{figure}[t]
\centering
\includegraphics[width=1\linewidth]{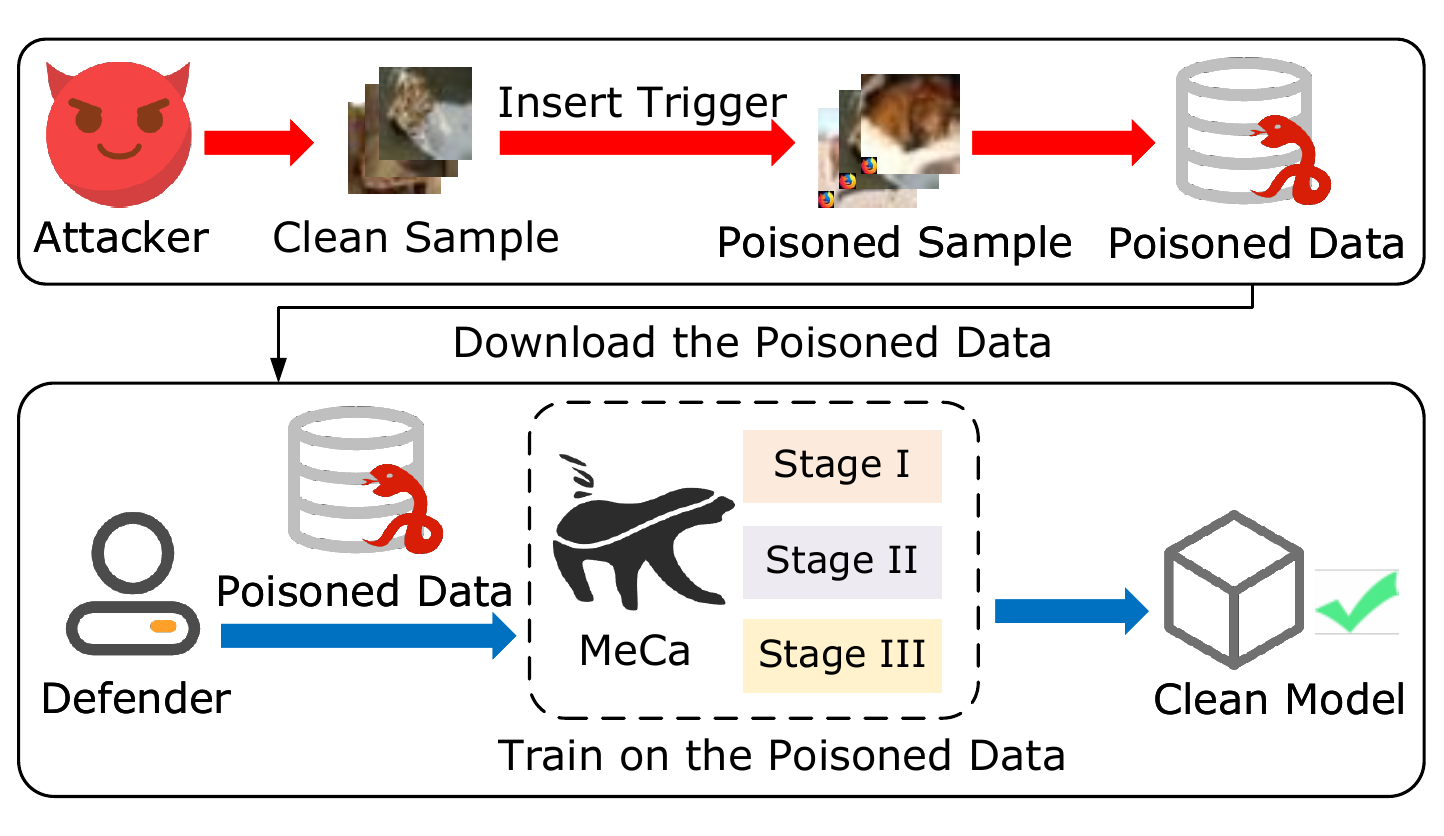}
\caption{Threat Model of \texttt{MeCa}.}
\label{fig:Threat Model}
\end{figure}

\subsection{Threat Model}
In the scenario considered in \texttt{MeCa}, the model trainer (Defender) intends to train a backdoor-robust model on the untrusted training data obtained from a third party. A malicious data provider (Attacker) endeavors to manipulate the dataset in order to induce the defender to train a model with a backdoor, thereby facilitating subsequent hijacking (e.g., targeted misclassification or creating authentication bypasses) of the model. The threat model is depicted in Fig.\ref{fig:Threat Model}. Therefore, in this paper, the defender utilizes \texttt{MeCa} to train a robust model on the untrusted dataset, free from the attacker's knowledge (e.g., triggers or target class). Considering the deployment and development of \texttt{MeCa} in real-world scenarios, we provide a detailed description of the goals, knowledge, and capabilities of both the Defender and the Attacker in the following.

\textbf{Defender's goal:} 
The defender seeks to detect poisoned samples and train a robust model from untrusted data, maintaining high task accuracy while resisting trigger-based attacks.

\textbf{Defender's knowledge and capabilities:} The defender defines the model architecture, learning algorithm, and hyperparameters and trains the model. It has no knowledge about the poison ratio, and without any auxiliary clean dataset, which is the recognized practice in real-world scenarios~\cite{pan2023asset} since poisoned samples can hardly be distinguished even for human experts.

\textbf{Attacker's goal:} 
The attacker compromises the model by injecting malicious samples into the training dataset to establish a backdoor.

\textbf{Attacker's knowledge and capabilities:} The attacker can manipulate the training dataset, including poisoning a fraction of the training dataset and modifying the labels of the poisoned samples. However, the attacker can not access the model architecture and parameters. Moreover, it can not interfere with the model training process.

\begin{figure*}
\centering
\includegraphics[scale=0.5]{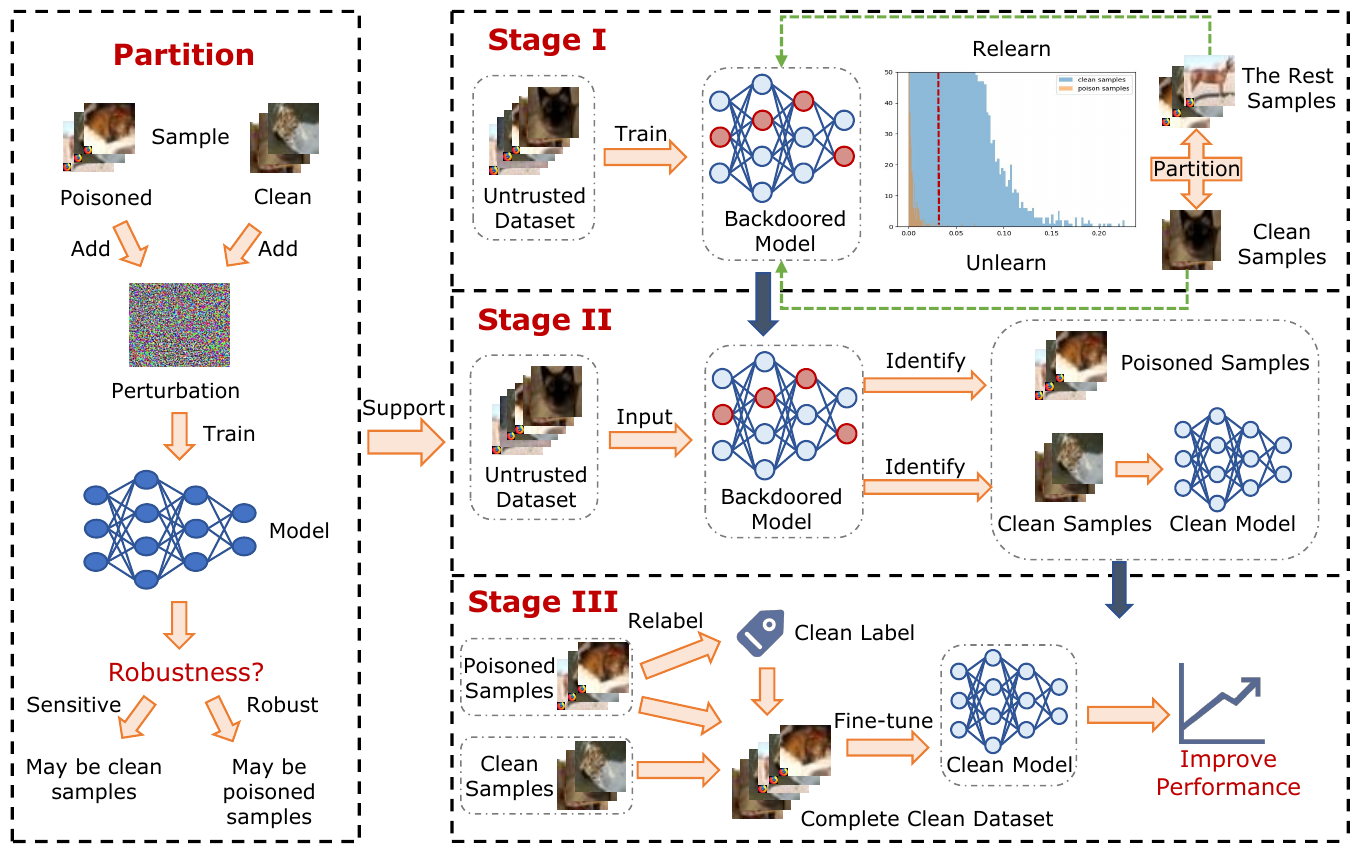}
\caption{Framework of \texttt{MeCa}.}
\label{fig:framework}
\end{figure*}

\section{Design of \texttt{MeCa}}
In this section, we provide an overview of the proposed method and the detailed design. 

\subsection{Overview of \texttt{MeCa}}

In this paper, we focus on how to identify the poisoned samples and train a clean model on the poisoned dataset. It is not trivial to satisfy this goal since even a tiny number of poisoned samples can accomplish the backdoor injection process. To address this problem, we first explore the inherent relationship between the perturbation and the backdoor attack theoretically and experimentally. Based on the results of the investigation, we propose a three-stage backdoor defense method dubbed \texttt{MeCa} and the overall framework is shown in Fig.\ref{fig:framework}. There are three main phases in the proposed \texttt{MeCa}. Firstly, we partition a tiny number of clean samples using the difference between clean samples and poisoned samples when subjected to perturbations. An enhanced backdoor model is trained by unlearning the chosen clean samples and relearning the remaining dataset. Then,  we accurately identify the poisoned and clean samples with a high degree of precision by leveraging the enhanced backdoor model. Specifically, samples that are misclassified by the enhanced backdoor model are deemed clean, while the remaining samples are classified as poisoned. Subsequently, a clean model can be trained using the clean samples identified. Finally, we relabel the poisoned samples and merge them with the clean ones to obtain a thoroughly clean dataset, enabling the fine-tuning of the clean model for enhanced performance. The design of \texttt{MeCa} is presented in detail as follows.

\subsection{Detailed Design of \texttt{MeCa}}
In this section, we present the detailed design of our backdoor defense approach.
\subsubsection{Exploring the Relationship between Perturbation and Backdoor}

To accurately identify poisoned samples, we need to enhance the backdoor model by unlearning certain clean samples. It is crucial to select a subset of clean samples without relying on an auxiliary dataset and to understand how poisoned samples affect the predictions made by backdoor models. Based on this understanding, we can design metrics to distinguish between clean and poisoned samples. 

Our investigation begins with a review of the backdoor attack process, which is similar with \cite{guo2022scale}. Let $\mathcal{D} = \{(\boldsymbol{x}_i, y_i)\}_{i=1}^N$ represent a clean training set, and let $f\in\mathcal{F}: \mathcal{X} \rightarrow \mathcal{Y}$ denotes the functionality of the target neural network. Each image $\boldsymbol{x}_i$ in $\mathcal{D}$ satisfies $\boldsymbol{x}_i \in \mathcal{X} = [0,1]^{C \times W \times H}$, and its corresponding label $y_i \in \mathcal{Y} = \{1, \ldots, K\}$, where $K$ is the total number of classes. To initiate an attack, adversaries poison a selection of clean samples $\mathcal{D}_p$ using a covert transformation $T(\cdot)$. These poisoned samples are then combined with the clean dataset prior to training the compromised model. This process can be formalized as $\mathcal{D}_t = \mathcal{D} \cup \mathcal{D}_p$, where $\mathcal{D}_p = \{(x_i^\prime, y_t) | x_i^\prime = T(x_i), (x_i, y_i) \in \mathcal{D}_p\}$. Various methods have been devised to make $T(\cdot)$ more subtle and harder to detect, or to reduce the assumptions about the adversary's capabilities, such as the poison ratio $pr = |\mathcal{D}_p| / |\mathcal{D}|$. Despite these variations, the ultimate goal remains the same \cite{guo2022scale}: manipulating the neural network by training a malicious model:
\begin{equation}
 \min _{\boldsymbol{\theta}} \sum_{i=1}^{N_b} \mathcal{L}\left(f\left(\boldsymbol{x}_i ; \boldsymbol{\theta}\right), y_i\right)+\sum_{j=1}^{N_p} \mathcal{L}\left(f\left(\boldsymbol{x}_j^{\prime} ; \boldsymbol{\theta}\right), y_t\right).
\end{equation}
where $N_b = |\mathcal{D}| $ and $N_p = |\mathcal{D}_p|$ and $\mathcal{L}$ typically stands for cross entropy.

\begin{figure*}[htp]
 \centering
 \subfloat[BadNets Attack]{\includegraphics[width=0.45\linewidth]{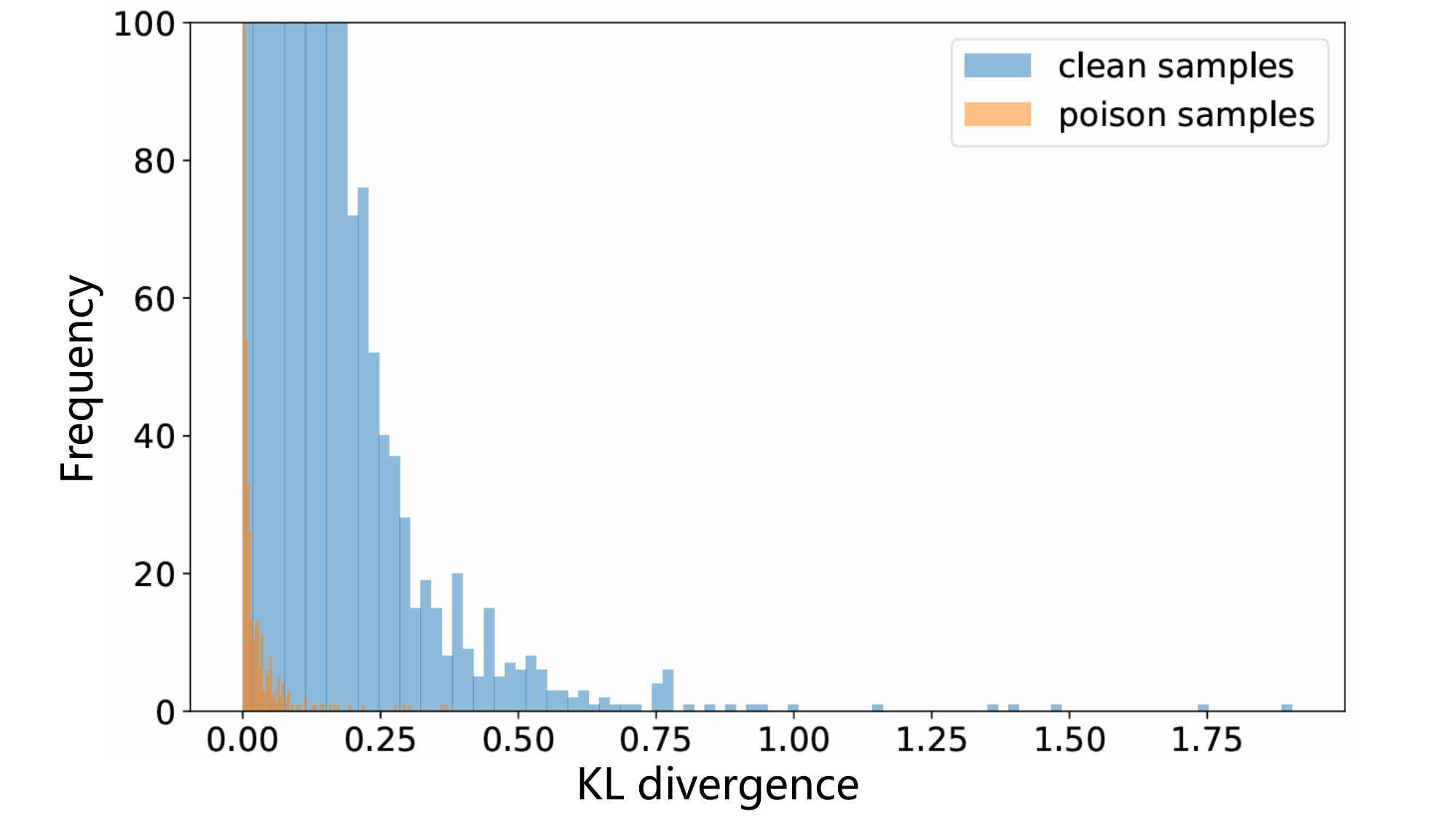}}
 \subfloat[Blend Attack]{\includegraphics[width=0.45\linewidth]{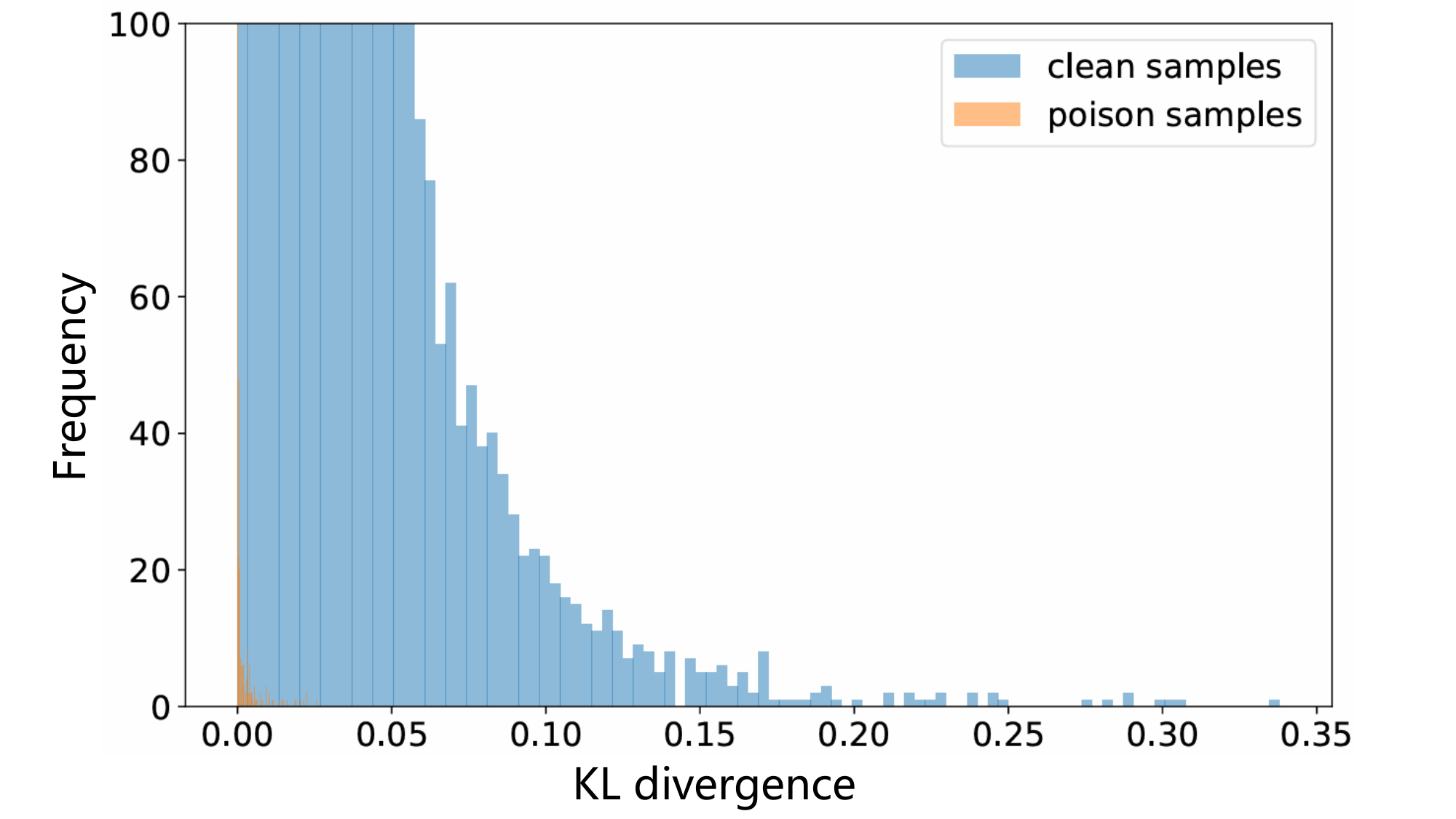}}
 \caption{Statistical histogram of training samples (with 5\% poison ratio on CIFAR10 using ResNet18).}
 \label{fig:histogram}
\end{figure*}

The trained malicious models behave abnormally on the poisoned samples while performing normally on the clean samples. The incentive behind such abnormal behavior is the cornerstone for designing a defense method against backdoor attacks. Khaddaj et al.\cite{khaddaj2023rethinking} validated that backdoor attacks corresponded to the strongest feature in the training data. Guo et al.\cite{guo2022scale} found that the predictions of poisoned samples were significantly more consistent compared to those of clean ones when amplifying all pixel values. The above observations indicate that the characteristics of the backdoor and clean ones differ, and it is possible to identify the backdoor samples. Inspired by the exploration of these previous works \cite{gao2023effectiveness, guo2022scale, ahmed2021threats}, we assume whether it's possible to identify poisoned samples by exploiting their robustness. To verify this, we conduct experiments on BadNets \cite{gu2019badnets} and Blend \cite{chen2017targeted} attacks on CIFAR10 with ResNet18. The poison ratio of these attacks is set to $5\%$ with a high attack success rate (ASR$\geq 99\%$). To measure the robustness of a sample, we adopt the metric measured by KL divergence below:
  \begin{equation}
 \mathcal{L}_{kl}(f(x), f(x\odot(1-\hat{m})+\hat{m}\odot\delta))).
  \end{equation}
where $\delta = \max(\min(\frac{\nabla_\delta \ell(f_\theta(x), y)}{\|\nabla_\delta \ell(f_\theta(x), y)\|_2}, \epsilon), -\epsilon)$ and $\hat{m}$ is a randomly-generated mask of the perturbation. As shown in Fig.\ref{fig:histogram}, on both BadNets and Blend attacks, most poisoned samples exhibit good consistency against adversarially generated perturbations $\delta$. 

To further explain this intriguing phenomenon, we propose the corresponding proof based on the recent studies \cite{guo2022scale} to analyze the characteristics of poisoned samples. We start with the regression solution for the Neural Tangent Kernel (NTK) model. The NTK is a theoretical framework that describes the training dynamics of deep neural networks in the limit of infinite width~\cite{jacot2018neural}. It establishes that for an infinitely wide network trained with gradient descent, the learning process simplifies to a kernel regression problem where the kernel remains constant throughout training. The NTK can also be used to formally analyze the characteristics of poisoned samples as presented in \cite{guo2021aeva,guo2022scale,guo2023scale}

\begin{equation}
\phi_t(\cdot) = \frac{\sum_{i=1}^{N_b} Q(\cdot, \boldsymbol{x}_i) \cdot y_i + \sum_{i=1}^{N_p} Q(\cdot, \boldsymbol{x}_i^{\prime}) \cdot y_t}{\sum_{i=1}^{N_b} Q(\cdot, \boldsymbol{x}_i) + \sum_{i=1}^{N_p} Q(\cdot, \boldsymbol{x}_i^{\prime})}
\end{equation}

Here:
\begin{itemize}
 \item $\phi_t(\cdot)$ stands for the predictive probability output of the function $f(\cdot; \theta)$ for class $t$,
 \item $\boldsymbol{x}_i$ are clean training samples,
 \item $y_i$ are the corresponding one-hot labels (so $y_i = 1$ if $\boldsymbol{x}_i$ belongs to class $t$, and $y_i = 0$ otherwise),
 \item $\boldsymbol{x}_i^{\prime}$ are the poisoned samples,
 \item $Q(\boldsymbol{x}, \boldsymbol{z}) = e^{-2 \gamma \|\boldsymbol{x} - \boldsymbol{z}\|^2}$ is the kernel function with $\gamma > 0$,
 \item $N_b$ and $N_p$ denote the number of clean and poisoned samples, respectively.
\end{itemize}

Assume the target label $y_t = 1$ and all other labels are 0. Because the training samples are uniformly distributed, $\frac{N_b}{k}$ clean samples are labeled as $y_t$. This allows us to simplify the equation to:

\begin{equation}
\phi_t(\cdot) = \frac{\sum_{i=1}^{N_b/k} Q(\cdot, \boldsymbol{x}_i) + \sum_{i=1}^{N_p} Q(\cdot, \boldsymbol{x}_i^{\prime})}{\sum_{i=1}^{N_b} Q(\cdot, \boldsymbol{x}_i) + \sum_{i=1}^{N_p} Q(\cdot, \boldsymbol{x}_i^{\prime})}
\end{equation}

Given that the backdoor sample $\boldsymbol{x}^{\prime}$ generally does not belong to the target class $y_t$, the contribution of clean samples in the numerator is negligible compared to the poisoned samples. Therefore, we can approximate $\phi_t(\cdot)$ by focusing primarily on the poisoned samples:

\begin{equation}
\phi_t(\cdot) \geq \frac{\sum_{i=1}^{N_p} Q(\cdot, \boldsymbol{x}_i^{\prime})}{\sum_{i=1}^{N_b} Q(\cdot, \boldsymbol{x}_i) + \sum_{i=1}^{N_p} Q(\cdot, \boldsymbol{x}_i^{\prime})}
\end{equation}

Consider a specific backdoor sample $\boldsymbol{x}^{\prime}$, which is crafted as a mixture of a clean sample $\boldsymbol{x}$ and a target pattern $\boldsymbol{t}$. When $N_p$ approaches $N_b$ (i.e., the poison ratio is close to 50\%), the attack's efficacy is maximized. In this case, the expression is:

\begin{equation}
\phi_t(\boldsymbol{x}^{\prime} + \boldsymbol{\delta}^{\prime}) \geq \frac{\sum_{i=1}^{N_p} Q\left(\boldsymbol{x}^{\prime}+\boldsymbol{\delta}^{\prime}, \boldsymbol{x}_{\boldsymbol{i}}^{\prime}\right)}{\sum_{i=1}^{N_b} Q\left(\boldsymbol{x}^{\prime}+\boldsymbol{\delta}^{\prime}, \boldsymbol{x}_{\boldsymbol{i}}\right)+\sum_{i=1}^{N_p} Q\left(\boldsymbol{x}^{\prime}+\boldsymbol{\delta}^{\prime}, \boldsymbol{x}_{\boldsymbol{i}}^{\prime}\right)}
\end{equation}

To satisfy the inequality below:

\begin{equation}
\phi_t(\boldsymbol{x}^{\prime} + \boldsymbol{\delta}^{\prime}) \geq \frac{1}{2},
\end{equation}
\textbf{a sufficient condition is that the sum of kernels for clean samples is less than or equal to that of poisoned samples}:

\begin{equation}
\sum_{i=1}^{N_b} Q\left(\boldsymbol{x}^{\prime}+\boldsymbol{\delta}^{\prime}, \boldsymbol{x}_{\boldsymbol{i}}\right)\leq\sum_{i=1}^{N_p} Q\left(\boldsymbol{x}^{\prime}+\boldsymbol{\delta}^{\prime}, \boldsymbol{x}_{\boldsymbol{i}}^{\prime}\right).
\end{equation}

Finally, using the kernel function $Q(\boldsymbol{x}, \boldsymbol{z}) = e^{-2\gamma \|\boldsymbol{x} - \boldsymbol{z}\|^2}$, we see that:

\begin{equation}
\sum_{i=1}^{N_p} \left(1 - e^{-2\gamma \|m \odot (\boldsymbol{t} - \boldsymbol{x}_i)\|^2}\right) > 0
\end{equation}

This inequality shows that the internal term is positive, implying that $f(\boldsymbol{x}^{\prime} + \boldsymbol{\delta}^{\prime}) = y_t$. Thus, the backdoor attack successfully causes the model to misclassify the poisoned sample as belonging to the target class $y_t$, confirming the attack's effectiveness. Based on the above proof, we can further obtain the theorem below:

\begin{figure}
 \centering
 \includegraphics[width=1\linewidth]{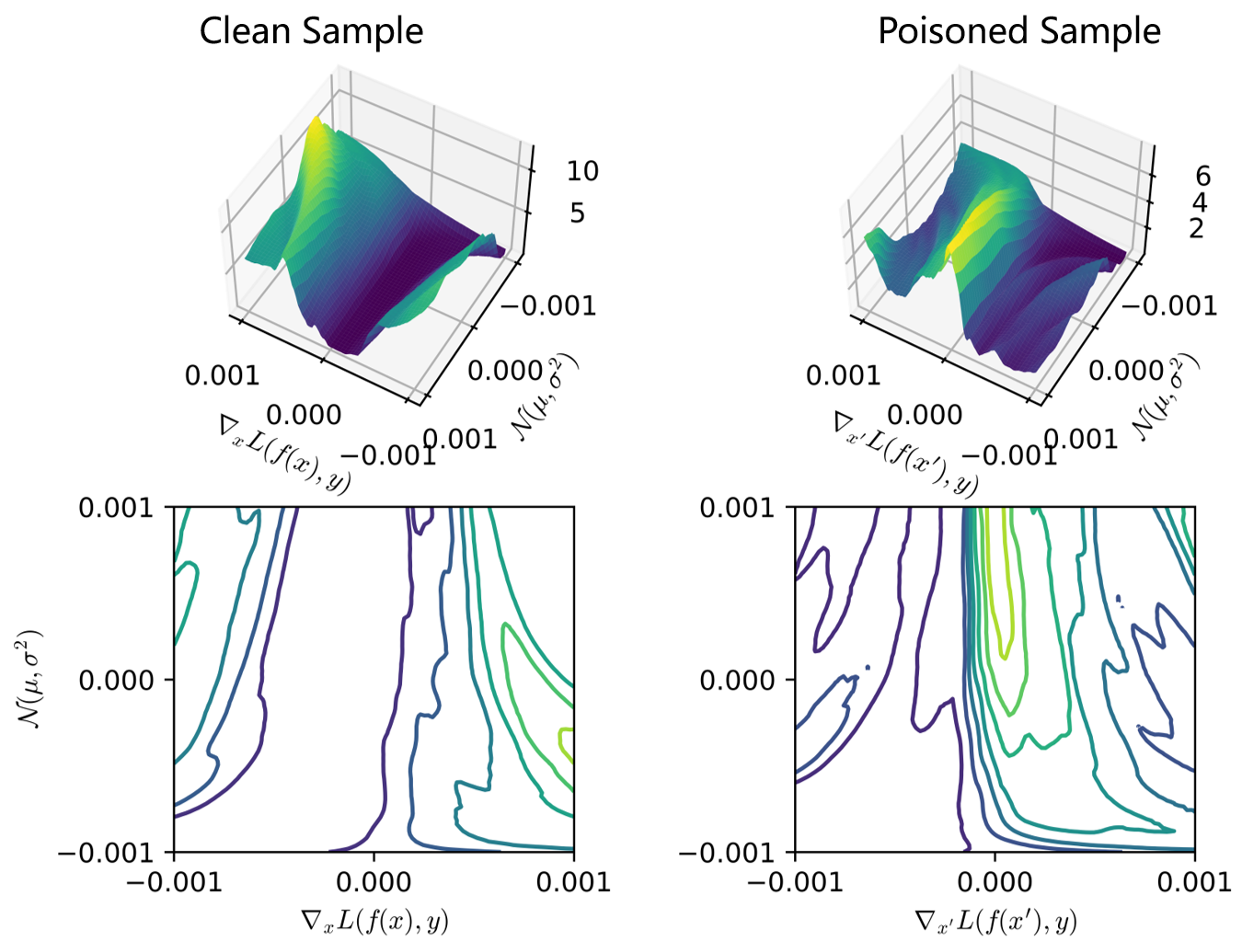}
 \caption{Loss landscapes for a representative clean and poisoned sample.}
 \label{fig:landscape}
\end{figure}

\textit{\textbf{Theorem 1}. Let $\mathcal{D} = \{(\boldsymbol{x}_i, y_i)\}_{i=1}^{N_b}$ be a clean dataset consisting of $N_b$ samples, and $\mathcal{D}_p = \{(\boldsymbol{x}_i', y_t')\}_{i=1}^{N_p}$ be a poisoned dataset containing $N_p$ samples, both independently and identically distributed (i.i.d.) according to a uniform distribution across $K$ classes. Suppose a deep neural network $f(\cdot; \theta)$ is formulated as a multivariate kernel regression using a Radial Basis Function (RBF) kernel, and shares the same objective as the attackers. For a given attacked sample $\boldsymbol{x}' = (\mathbf{1} - \boldsymbol{m}) \odot \boldsymbol{x} + \boldsymbol{m} \odot \boldsymbol{t}$, where $\mathbf{1}$ is a vector of ones, $\boldsymbol{m}$ is a binary mask, $\boldsymbol{x}$ is the original sample, and $\boldsymbol{t}$ is the trigger pattern, we have: $\lim_{N_p \to N_b} C(\boldsymbol{x}' + \delta) = y_t$, where $C(\cdot)$ denotes the classification function of the network, $\delta$ represents a perturbation added to the attacked sample, and $y_t$ is the target class label.}

The theorem above demonstrates that when the size of $\mathcal{D}_p$ approaches that of the clean dataset, poisoned samples exhibit robustness against potential perturbations. Note that we must acknowledge there is an apparent gap between the theoretical assumption in Theorem 1, which considers an asymptotic case where $N_p\rightarrow N_b$ (poison ratio approaches 50\%), and the lower, more practical ratios in our experiments. Our intention with the theorem was to provide a theoretical intuition in an analytically tractable setting to explain why backdoor features become exceptionally robust. The theorem demonstrates that under these conditions, the backdoor becomes the dominant feature. Our extensive empirical results then demonstrate that this principle of differential robustness holds true and is practically effective even at much lower poisoning rates.

To gain a further intuitive understanding, we visualize the loss landscapes of both clean and poisoned samples (using the Blend attack~\cite{chen2017targeted}), as illustrated in Fig.~\ref{fig:landscape}. Notably, the loss of a clean sample shows steeper changes when perturbations are applied in the direction of the gradient. This observation further corroborates the previous theoretical analysis~\cite{zeng2024clibe} in which the author found that the square sum of the eigenvalues of the Hessian matrix of the benign model is higher than that of the backdoored one. To further enlarge such a sensitivity (robustness) gap, we apply the worst-case perturbations (adversarial perturbation) to make poisoned samples as distinguishable as possible.

\begin{algorithm}
 \caption{Untrusted Dataset Partition}
 \label{agm:partition}
 \begin{algorithmic}[1]  
 \REQUIRE Untrusted Dataset $\mathcal{D}_u$, Backdoored Model $\hat{f}$, Perturbation Radius $\epsilon$, Patch Size $r$, Partition Rate $p$
 \ENSURE  Partitioned Dataset $\mathcal{D}_u$ and $\mathcal{D}_c$
 \STATE $Q=[\  ]$  // \textit{Initialize empty list $Q$}
	\FOR {$(x,y)$\  in\  $\mathcal{D}_u$}
  \STATE // \textit{Generate a mask of the perturbation patch}
  \STATE $\hat{m} = RandomPatchMask(r)$
  \STATE $\delta = \max(\min(\frac{\nabla_\delta \ell(f_\theta(x), y)}{\|\nabla_\delta \ell(f_\theta(x), y)\|_2}, \epsilon), -\epsilon)$\\
 \STATE $\hat{x}=x\odot(1-\hat{m})+\hat{m}\odot\delta$
 \STATE // \textit{Calculate $KL$ divergence of each sample}
 \STATE $Q.append(\mathcal{L}_{kl}(f(x), f(\hat{x})))$
 \ENDFOR
\STATE // \textit{Partition $\mathcal{D}_u$ according to $Q$ and $p$}
\STATE $\mathcal{D}_u,\mathcal{D}_c = sort(\mathcal{D}_u,Q,p)$
\RETURN $\mathcal{D}_u,\mathcal{D}_c$
\end{algorithmic}
\end{algorithm}

\begin{figure}
 \centering
 \includegraphics[width=0.95\linewidth]{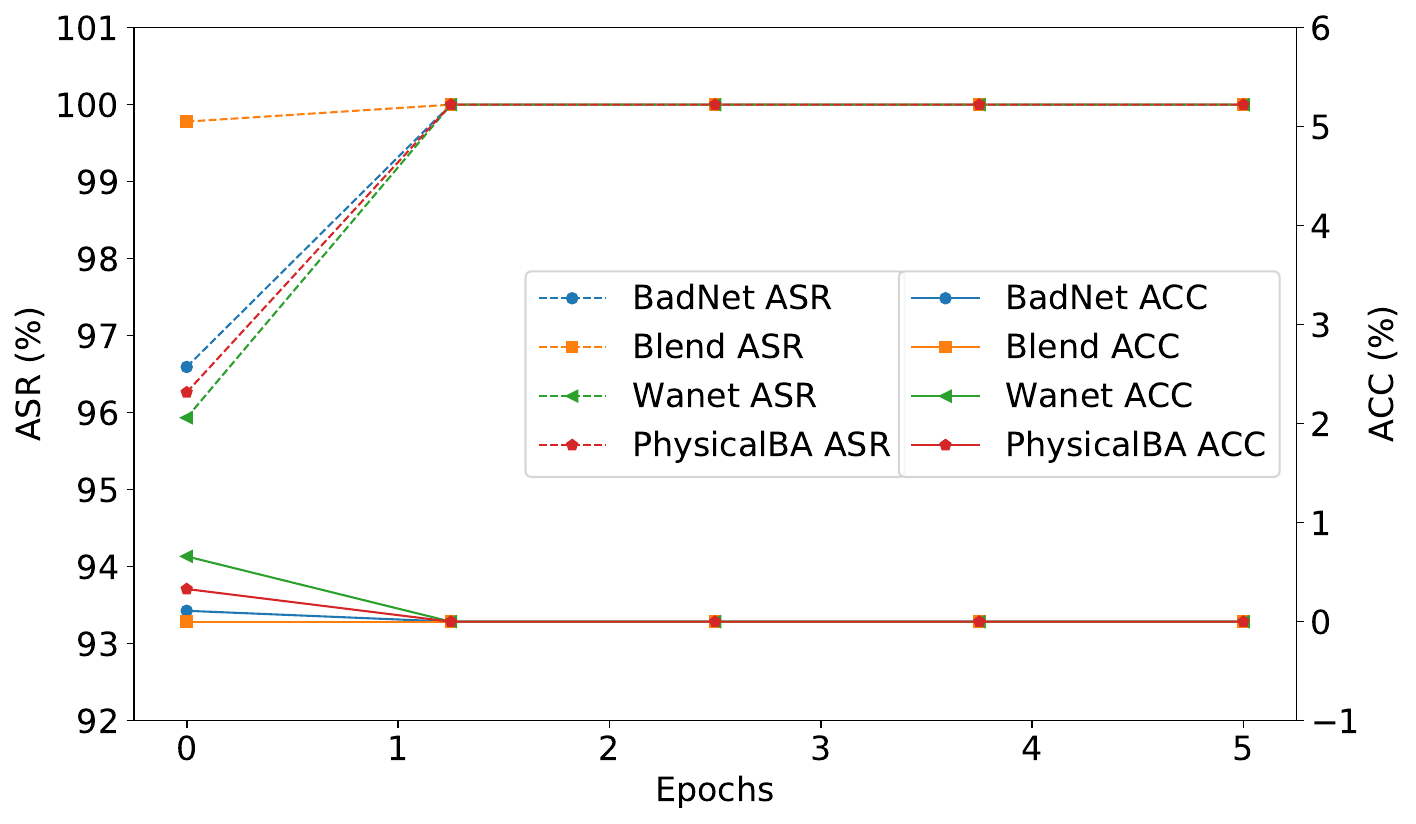}
 \caption{ASR and ACC of the model during backdoor-enhanced training on CIFAR10 with ResNet18.}
 \label{fig:backdoor-enhancement}
\end{figure}

\subsubsection{Dataset Partition}
Based on the findings in the above section, we plan to partition the untrusted dataset roughly by using the characteristic that the poisoned samples perform greater robustness against potential perturbations than the clean samples. However, as shown in Fig.\ref{fig:histogram}, these two histograms (y-axis means number and x-axis means KL divergence) illustrate that poisoned samples are still mixed with some clean ones. Therefore, in this section, we only pick out a small number of clean samples iteratively during the training stage rather than precisely selecting all the poisoned samples. 

Algorithm \ref{agm:partition} shows the complete process of our dataset partition operation. This algorithm delineates a meticulous process for partitioning a dataset into trusted and potentially compromised subsets by exploiting the variances in the behavior of a model when subjected to perturbed inputs. At its core, the algorithm iterates over each sample $(x,y)$ in the dataset $\mathcal{D}_u$, applying a strategically-sized perturbation patch characterized by a radius $r$ and bounded by a perturbation radius $\epsilon$, to the input sample. This is achieved through generating a random patch mask $\hat{m}$ and calculating an optimal adversarial perturbation $\delta$, which is applied to the input $x$ to yield a perturbed counterpart $\hat{x}$. The key metric for partitioning, the KL divergence, is computed between the model's predictions on the original and perturbed inputs, with these divergence scores being aggregated into a list $Q$. Subsequently, the dataset is partitioned into two subsets, $\mathcal{D}_u$ and $\mathcal{D}_c$, based on the divergence scores relative to a partition rate $p$, indexing subsets with assumed clean and those potentially poisoned, respectively. This method essentially leverages the sensitivity of backdoored models to perturbations in identifying and isolating suspicious data samples, thus serving as a robust strategy in the preprocessing phase of data handling. With the above steps, we obtain $\mathcal{D}_u$ and $\mathcal{D}_c$. Here, $\mathcal{D}_c$ denotes the selected small number of clean samples. $\mathcal{D}_u$ represents the rest of the training data that contains the poisoned samples.

\subsubsection{Backdoor Enhancement and Standard Training}
Note that a small number of clean samples $\mathcal{D}_c$ means nothing to train a clean model. However, motivated by \cite{li2021anti}, a small number of poisoned samples is enough to unlearn the functionality of the backdoor. We instead unlearn the main task on the ``possibly'' backdoored model with selected benign samples. That's to say, we can obtain a model $\hat{f}_{\theta}$ with a high ASR but low ACC by unlearning some clean samples, which can be recognized as a poisoned sample detector. To achieve a higher ASR and lower ACC, we train a backdoor-enhanced model iteratively by unlearning a small number of clean samples and learning the rest poisoned dataset. Concretely, for each iteration, $p$ of samples with the lowest consistency will be labeled as clean samples $\mathcal{D}_c$ and unlearned:
\begin{equation}
 \theta = {\arg\max\limits_{\theta}}_{(x,y)\in\mathcal{D}_c} l(\hat{f}_{\theta}(x),y).
\end{equation}

During this process, more and more clean samples will be picked out for unlearning with the decrease of $p$, and the model will also be trained on a dataset with an increasing poison ratio. The rationale for designing for adopting such a dynamic process is that the $p$ must be small for its weak ability to detect poisoned samples. While unlearning with more benign samples and learning with more poisoned samples, the ability of the detection increases as well and a higher $p$ is set to pick out more benign samples. As illustrated in Fig.\ref{fig:backdoor-enhancement}, ASR and ACC of the model continue to increase and decline, respectively. Algorithm \ref{agm:standard} shows the iterative backdoor-enhanced training process. 

\begin{algorithm}
 \caption{Backdoor Enhancement and Standard Training}
 \label{agm:standard}
 \begin{algorithmic}[1]  
 \REQUIRE Untrusted Dataset $\mathcal{D}_u$, Backdoor Training Epochs $E_b$, Standard Training Epochs $E_s$
 \ENSURE  Clean Model $f_{\theta}$
\STATE $\mathcal{\hat{D}}_u = \mathcal{D}_u$  // \textit{Initialize $\mathcal{\hat{D}}_u$}
	\FOR {$e$\  in\  $range(E_b)$}
\STATE $\theta = {\arg\min\limits_{\theta}}_{(x,y)\in\mathcal{\hat{D}}_u} l(\hat{f}_{\theta}(x),y) $
 \STATE // \textit{Update $\mathcal{\hat{D}}_u$ with partition algorithm}
 \STATE // \textit{$\gamma$ is a coefficient}
 \STATE $p=1-(e+1)\times\gamma$  
 \STATE $\mathcal{\hat{D}}_u,\mathcal{D}_c = Partition(\mathcal{D}_u,\hat{f}_{\theta},p)$
 \STATE // \textit{Unlearn the clean samples}
 \STATE $\theta = {\arg\max\limits_{\theta}}_{(x,y)\in\mathcal{D}_c} l(\hat{f}_{\theta}(x),y) $
 \ENDFOR
\STATE // \textit{Pick out the clean samples}
\STATE$\mathcal{D}_c = \{ (x,y) \in \mathcal{D}_u : \hat{f}_{\theta}(x)\neq y \}$
 \FOR {$e$\  in\  $range(E_s)$}
\STATE $\theta = {\arg\min\limits_{\theta}}_{(x,y)\in\mathcal{D}_c} l(f_{\theta}(x),y) $
 \ENDFOR
\RETURN $f_\theta$
\end{algorithmic}
\end{algorithm}

\begin{table*}[h]\centering
\caption{The detailed settings of the $8$ state-of-the-art backdoor attacks.}
\label{Table:attacks}
\renewcommand{\arraystretch}{1.5}
\begin{tabular}{ccccccccc}
\hline
Attack Method & Poison Ratio & Trigger & Static/Dynamic & Dirty/Clean & Adaptive & Source & Target & Cover Rate \\ \hline
\rowcolor[HTML]{EFEFEF} 
BadNet & 5\%  & 3*3 white square& Static  & Dirty& No& / & 1 & /\\
Blend  & 5\%  & HelloKity , alpha=0.2  & Static  & Dirty& No& / & 1 & /\\
\rowcolor[HTML]{EFEFEF} 
SIG & 5\%  & Sinusoidal signal & Static  & Clean& No& / & 1 & /\\
WaNet  & 5\%  & Warping-based triggers & Dynamic & Dirty& No& / & 1 & 5\%\\
\rowcolor[HTML]{EFEFEF} 
PhysicalBA & 5\%  & Firefox & Static  & Dirty& No& / & 1 & /\\
TaCT& 5\%  & 3*3 white square& Static  & Dirty& Yes & 0 & 1 & 0.01\\
\rowcolor[HTML]{EFEFEF} 
AdaptiveBlend & 5\%  & HelloKity , alpha=0.2  & Static  & Dirty& Yes & / & 1 & 0.01\\
AdaptivePatch & 5\%  & Firefox*4 & Static  & Dirty& Yes & / & 1 & 0.01\\ \hline
\end{tabular}
\end{table*}

Obviously, our method does not end with training a backdoor-enhanced model, which plays a key role in identifying the poisoned samples. It's natural to figure out the misclassified samples on the backdoored model and mark them as clean samples while the rest are labeled as poisoned samples. Finally, with the selected clean samples, we can train a clean model.

\subsubsection{Relabel and Relearn}
To obtain a completed clean dataset and further improve the performance of the clean model, \texttt{MeCa} relabels the poisoned samples and merges them with clean samples to obtain a clean and complete dataset: 
\begin{equation}
 \mathcal{D}_{com} = \mathcal{D}_{clean}\cup\{(x_i,\hat{y}_i)|\hat{y}_i=f(x),(x_i,y_i)\in\mathcal{D}_p\}.
\end{equation}
where $\mathcal{D}_{clean}$ denotes the selected clean samples based on the backdoor-enhanced model and $\mathcal{D}_{com}$ represents the complete clean dataset. Then, the complete dataset can be used to fine-tune the clean model to further improve its performance. Note that our experiment mainly shows the results before relabeling and relearning to follow the setting in previous works \cite{li2021anti,chen2022effective}, and the performance of the model fine-tuned on the complete dataset is given separately.

\section{Experiments}

In this section, we experimentally evaluate the performance of the proposed \texttt{MeCa}. Firstly, we show the experimental settings, including the experimental environment, datasets, networks, backdoor attacks, and backdoor defenses. Then, we analyze and summarize the experimental results. Finally, we also present the ablation studies.  
\subsection{Experimental Settings}

\subsubsection{Datatsets}

We run our experiments on four image datasets, including Imagnette, Tiny ImageNet, CIFAR10, and CIFAR100. The introduction of the above four datasets is shown as follows.
\begin{itemize}
 \item \textbf{Imagnette} \cite{howard2020fastai}: Imagnette is a small dataset extracted from the large dataset ImageNet (more than $14$ million images, $20,000$ categories). Imagenette's training set contains $9,469$ images, and its test set contains $3,925$ images, all in JPEG format. The image resolution is not uniform, but the width and height are no less than $160$ pixels. 
 \item \textbf{Tiny ImageNet} \cite{le2015tiny}: Tiny ImageNet contains $100,000$ images of $200$ classes ($500$ for each class) downsized to $64 \times 64$ colored images. Each class has $500$ training images, $50$ validation images, and $50$ test images.
 \item \textbf{CIFAR10}\cite{krizhevsky2009learning}: CIFAR10 dataset is a database of tiny images, containing $50,000$ training images and $10,000$ testing images with $10$ classes.
 \item \textbf{CIFAR100} \cite{krizhevsky2009learning}: CIFAR100 contains $100$ categories. Each category contains $500$ training images with $32 \times 32$ and $100$ test images with $32 \times 32$. 
\end{itemize}

\subsubsection{Networks}

We conduct our experiments on three deep learning models: ResNet18, ResNet34, and MobileNetV2. Unless otherwise stated, the default experimental setup assumes the use of ResNet18.

\subsubsection{Experimental Environment}
All experiments were conducted on a server equipped with Intel i9-9900K, 3.60GHz processor, 32GB RAM, NVIDIA GeForce RTX 3090, and PyTorch.

\subsubsection{Backdoor Attacks} We employ $8$ state-of-the-art backdoor attacks to evaluate the proposed \texttt{MeCa} in our experiments, including $3$ dirty-label attacks: BadNet \cite{gu2019badnets}, Blend \cite{chen2017targeted}, and PhysicalBA \cite{li2021backdoor}, one clean-label attack: SIG \cite{barni2019new}, one input-specific-trigger attack: WaNet \cite{nguyen2021wanet}, and $3$ adaptive attacks: TaCT \cite{tang2021demon}, AdaptiveBlend \cite{qi2022revisiting} and AdaptivePatch \cite{qi2022revisiting}. The poison ratio is set to $5\%$, aligning with most previous works~\cite{li2021anti,tang2021demon,chen2022effective,liu2018fine,borgnia2021strong,zheng2022data}. The detailed configurations of these backdoor attacks are shown in Table \ref{Table:attacks}.

\subsubsection{Backdoor Defense}
We compare \texttt{MeCa} with $7$ established backdoor defense baselines, including FineTuning \cite{liu2018fine}, FinePruning \cite{liu2018fine}, CutMix \cite{borgnia2021strong}, CLP \cite{zheng2022data}, DBR \cite{chen2022effective}, SCAnFT \cite{tang2021demon}, and ABL \cite{li2021anti}.

Our detailed configurations for baseline defense (CIFAR10) are listed as follows:
\begin{itemize}
 \item \textbf{FineTuning:} We use $2,000$ reserved clean samples to finetune the ``full layer'' of the model, and the learning rate in the repair phase is set to $0.001$ with $10$ epochs.
 \item \textbf{FinePruning:} We use $2,000$ reserved clean samples to prune the $30\%$ of the channels of the second layer.
 \item \textbf{CutMix:} Probability of CutMix is set to $1$ with $\beta=1,\gamma=1$. To repair the backdoored model, the learning rate in the repair phase is $0.01$ for $10$ epochs.
 \item \textbf{CLP:} Similarly, CLP needs to prune abnormal channels of the model that are out of $3$ standard deviations and repair the model for $10$ epochs.
 \item \textbf{DBR:} DBR tries to pick out with high sensitivity to ``rotate'' and ``affine'' transformation. We set clean ratio and poison ratio to $40\%$ and $5\%$, respectively, and unlearn the poisoned samples for $5$ epochs with $0.0001$ unlearning rate.
 \item \textbf{SCAnFT:} SCAnFT is a poisoned sample detection algorithm and here we adopt it to pick out the triggered samples, and unlearn them with the unlearning rate set to $0.0001$.
 \item \textbf{ABL:} ABL tries to pick out poisoned samples by identifying the samples with lower loss values, and the threshold $\gamma$ of loss values is set to $0.8$, and the unlearning rate is $0.0001$ for $5$ epochs.
 \item \textbf{MeCa:} For our defense, we only apply a single-step backward to obtain the perturbation bounded by $0.001$, and the patch size is set to 2 for all attacks. Note that both of the constraints make it impossible to affect the functionality of the backdoor while slightly shifting the logits distribution of the clean samples. The unlearning coefficient is set to 0.05 across any datasets.
\end{itemize}

\subsubsection{Evaluation Metrics} We employ the ASR and ACC to evaluate the performance of the proposed backdoor defense framework. ASR indicates the ratio of the triggered samples that are misclassified as the target label. ACC represents the accuracy of the model on the clean samples. The lower ASR and the higher ACC indicate the better performance of the backdoor defense mechanism. 

Specifically, we suppose a defense is successful if the post-defense ASR is under $20\%$, and unsuccessful otherwise, as done in prior works \cite{qi2023towards, xie2023badexpert}.

\begin{table*}[htp]\centering
\caption{The comparison results (\%) between our \texttt{MeCa} and $7$ state-of-the-art backdoor defenses methods against $5$ state-of-the-art backdoor attacks on ResNet18 with $5\%$ poisoned ratio.}
\label{tab:main result}
\renewcommand{\arraystretch}{2}
\resizebox{2\columnwidth}{!}{
\begin{tabular}{cccccccccccccccccccc}
\hline
& & \multicolumn{2}{c}{No Defense} & \multicolumn{2}{c}{FineTuning}& \multicolumn{2}{c}{FinePruning}& \multicolumn{2}{c}{CutMix} & \multicolumn{2}{c}{CLP} & \multicolumn{2}{c}{DBR}& \multicolumn{2}{c}{SCAnFT}   & \multicolumn{2}{c}{ABL}& \multicolumn{2}{c}{\texttt{MeCa}} \\ \cline{2-20} 
\multirow{-2}{*}{Dataset} & Types   & ACC& ASR & ACC & ASR & ACC& ASR& ACC& ASR& ACC& ASR & ACC& ASR& ACC & ASR& ACC & ASR & ACC & ASR   \\ \hline
& \cellcolor[HTML]{EFEFEF}BadNet  & \cellcolor[HTML]{EFEFEF}94.10 & \cellcolor[HTML]{EFEFEF}93.70  & \cellcolor[HTML]{EFEFEF}71.47& \cellcolor[HTML]{EFEFEF}32.13  & \cellcolor[HTML]{EFEFEF}61.50 & \cellcolor[HTML]{EFEFEF}54.85 & \cellcolor[HTML]{EFEFEF}72.30 & \cellcolor[HTML]{EFEFEF}56.00 & \cellcolor[HTML]{EFEFEF}33.24 & \cellcolor[HTML]{EFEFEF}6.09   & \cellcolor[HTML]{EFEFEF}69.81 & \cellcolor[HTML]{EFEFEF}57.62 & \cellcolor[HTML]{EFEFEF}46.54& \cellcolor[HTML]{EFEFEF}16.34 & \cellcolor[HTML]{EFEFEF}57.61& \cellcolor[HTML]{EFEFEF}58.45  & \cellcolor[HTML]{EFEFEF}\textbf{73.15} & \cellcolor[HTML]{EFEFEF}\textbf{0.00} \\
& Blend   & 96.60 & 95.50  & 70.08  & 88.92  & 60.94 & 93.63 & 55.12 & 92.80 & 13.30 & 0.00& 45.15 & 70.64 & 57.89  & 17.17 & 62.60  & 79.22  & \textbf{73.13} & \textbf{0.00} \\
& \cellcolor[HTML]{EFEFEF}SIG & \cellcolor[HTML]{EFEFEF}96.40 & \cellcolor[HTML]{EFEFEF}86.30  & \cellcolor[HTML]{EFEFEF}70.36& \cellcolor[HTML]{EFEFEF}71.47  & \cellcolor[HTML]{EFEFEF}60.39 & \cellcolor[HTML]{EFEFEF}15.51 & \cellcolor[HTML]{EFEFEF}70.63 & \cellcolor[HTML]{EFEFEF}88.09 & \cellcolor[HTML]{EFEFEF}16.90 & \cellcolor[HTML]{EFEFEF}74.24  & \cellcolor[HTML]{EFEFEF}61.50 & \cellcolor[HTML]{EFEFEF}0.28  & \cellcolor[HTML]{EFEFEF}41.83& \cellcolor[HTML]{EFEFEF}0.28  & \cellcolor[HTML]{EFEFEF}16.90& \cellcolor[HTML]{EFEFEF}83.38  & \cellcolor[HTML]{EFEFEF}\textbf{73.96} & \cellcolor[HTML]{EFEFEF}\textbf{0.00} \\
& WaNet   & 94.30 & 45.90  & 69.81  & 11.08  & 59.56 & 41.00 & 68.98 & 49.86 & 29.36 & 48.48  & 64.82 & 7.20  & 52.35  & 12.47 & 36.84  & 78.67  & \textbf{70.64} & \textbf{0.00} \\
& \cellcolor[HTML]{EFEFEF}PhysicalBA & \cellcolor[HTML]{EFEFEF}89.30 & \cellcolor[HTML]{EFEFEF}100.00 & \cellcolor[HTML]{EFEFEF}58.73& \cellcolor[HTML]{EFEFEF}100.00 & \cellcolor[HTML]{EFEFEF}54.02 & \cellcolor[HTML]{EFEFEF}99.72 & \cellcolor[HTML]{EFEFEF}56.79 & \cellcolor[HTML]{EFEFEF}59.78 & \cellcolor[HTML]{EFEFEF}23.27 & \cellcolor[HTML]{EFEFEF}0.00   & \cellcolor[HTML]{EFEFEF}63.43 & \cellcolor[HTML]{EFEFEF}6.37  & \cellcolor[HTML]{EFEFEF}\textbf{77.03} & \cellcolor[HTML]{EFEFEF}10.77 & \cellcolor[HTML]{EFEFEF}63.37& \cellcolor[HTML]{EFEFEF}2.77   & \cellcolor[HTML]{EFEFEF}60.11& \cellcolor[HTML]{EFEFEF}\textbf{0.00} \\
\multirow{-6}{*}{Imagenette}   & Average & 94.14 & 84.28  & 68.09  & 60.72  & 59.28 & 60.94 & 64.76 & 69.31 & 23.21 & 25.76  & 60.94 & 28.42 & 55.13  & 11.41 & 47.46  & 60.49  & \textbf{70.20} & \textbf{0.00} \\ \hline
& \cellcolor[HTML]{EFEFEF}BadNet  & \cellcolor[HTML]{EFEFEF}92.00 & \cellcolor[HTML]{EFEFEF}95.20  & \cellcolor[HTML]{EFEFEF}83.30& \cellcolor[HTML]{EFEFEF}96.81  & \cellcolor[HTML]{EFEFEF}75.05 & \cellcolor[HTML]{EFEFEF}93.19 & \cellcolor[HTML]{EFEFEF}81.21 & \cellcolor[HTML]{EFEFEF}83.30 & \cellcolor[HTML]{EFEFEF}34.18 & \cellcolor[HTML]{EFEFEF}0.55   & \cellcolor[HTML]{EFEFEF}67.47 & \cellcolor[HTML]{EFEFEF}1.21  & \cellcolor[HTML]{EFEFEF}69.89& \cellcolor[HTML]{EFEFEF}2.09  & \cellcolor[HTML]{EFEFEF}71.98& \cellcolor[HTML]{EFEFEF}0.00   & \cellcolor[HTML]{EFEFEF}\textbf{85.72} & \cellcolor[HTML]{EFEFEF}\textbf{0.00} \\
& Blend   & 84.50 & 99.90  & 84.62  & 100.00 & 79.23 & 98.79 & 80.44 & 99.67 & 51.00 & 3.52& 37.03 & 98.46 & 78.57  & 1.43  & 82.20  & 5.71& \textbf{85.71} & \textbf{0.00} \\
& \cellcolor[HTML]{EFEFEF}SIG & \cellcolor[HTML]{EFEFEF}84.70 & \cellcolor[HTML]{EFEFEF}95.40  & \cellcolor[HTML]{EFEFEF}82.31& \cellcolor[HTML]{EFEFEF}95.93  & \cellcolor[HTML]{EFEFEF}80.88 & \cellcolor[HTML]{EFEFEF}90.11 & \cellcolor[HTML]{EFEFEF}79.12 & \cellcolor[HTML]{EFEFEF}99.12 & \cellcolor[HTML]{EFEFEF}37.47 & \cellcolor[HTML]{EFEFEF}71.97  & \cellcolor[HTML]{EFEFEF}38.57 & \cellcolor[HTML]{EFEFEF}93.52 & \cellcolor[HTML]{EFEFEF}66.92& \cellcolor[HTML]{EFEFEF}3.63  & \cellcolor[HTML]{EFEFEF}85.93& \cellcolor[HTML]{EFEFEF}48.35  & \cellcolor[HTML]{EFEFEF}\textbf{87.69} & \cellcolor[HTML]{EFEFEF}\textbf{2.64} \\
& WaNet   & 82.80 & 51.30  & 83.41  & 70.33  & 79.78 & 61.65 & 76.70 & 31.10 & 12.75 & 78.02  & 26.70 & 6.15  & 56.70  & 60.77 & \textbf{83.74} & 88.57  & 83.63  & \textbf{0.00} \\
& \cellcolor[HTML]{EFEFEF}PhysicalBA & \cellcolor[HTML]{EFEFEF}91.60 & \cellcolor[HTML]{EFEFEF}100.00 & \cellcolor[HTML]{EFEFEF}86.04& \cellcolor[HTML]{EFEFEF}100.00 & \cellcolor[HTML]{EFEFEF}80.11 & \cellcolor[HTML]{EFEFEF}49.78 & \cellcolor[HTML]{EFEFEF}85.38 & \cellcolor[HTML]{EFEFEF}79.23 & \cellcolor[HTML]{EFEFEF}27.47 & \cellcolor[HTML]{EFEFEF}0.00   & \cellcolor[HTML]{EFEFEF}30.55 & \cellcolor[HTML]{EFEFEF}5.71  & \cellcolor[HTML]{EFEFEF}79.01& \cellcolor[HTML]{EFEFEF}0.33  & \cellcolor[HTML]{EFEFEF}\textbf{87.14} & \cellcolor[HTML]{EFEFEF}0.99   & \cellcolor[HTML]{EFEFEF}82.97& \cellcolor[HTML]{EFEFEF}\textbf{0.00} \\
\multirow{-6}{*}{CIFAR10} & Average & 87.12 & 88.36  & 83.94  & 92.61  & 79.01 & 78.70 & 80.57 & 78.48 & 32.57 & 30.81  & 40.06 & 41.01 & 70.22  & 13.65 & 82.20  & 28.72  & \textbf{85.14} & \textbf{0.53} \\ \hline
& \cellcolor[HTML]{EFEFEF}BadNet  & \cellcolor[HTML]{EFEFEF}72.10 & \cellcolor[HTML]{EFEFEF}98.20  & \cellcolor[HTML]{EFEFEF}58.40& \cellcolor[HTML]{EFEFEF}91.40  & \cellcolor[HTML]{EFEFEF}42.00 & \cellcolor[HTML]{EFEFEF}91.09 & \cellcolor[HTML]{EFEFEF}42.91 & \cellcolor[HTML]{EFEFEF}86.03 & \cellcolor[HTML]{EFEFEF}54.55 & \cellcolor[HTML]{EFEFEF}83.30  & \cellcolor[HTML]{EFEFEF}29.35 & \cellcolor[HTML]{EFEFEF}0.10  & \cellcolor[HTML]{EFEFEF}56.78& \cellcolor[HTML]{EFEFEF}0.00  & \cellcolor[HTML]{EFEFEF}\textbf{63.16} & \cellcolor[HTML]{EFEFEF}0.10   & \cellcolor[HTML]{EFEFEF}60.53& \cellcolor[HTML]{EFEFEF}\textbf{0.00} \\
& Blend   & 51.90 & 100.00 & 60.12  & 99.80  & 31.58 & 99.79 & 48.99 & 99.70 & 34.31 & 59.51  & 8.20  & 89.88 & 48.89  & 0.00  & \textbf{61.40} & 1.21& 60.53  & \textbf{0.00} \\
& \cellcolor[HTML]{EFEFEF}SIG & \cellcolor[HTML]{EFEFEF}54.40 & \cellcolor[HTML]{EFEFEF}95.00  & \cellcolor[HTML]{EFEFEF}60.32& \cellcolor[HTML]{EFEFEF}96.86  & \cellcolor[HTML]{EFEFEF}40.28 & \cellcolor[HTML]{EFEFEF}67.71 & \cellcolor[HTML]{EFEFEF}48.99 & \cellcolor[HTML]{EFEFEF}95.95 & \cellcolor[HTML]{EFEFEF}43.72 & \cellcolor[HTML]{EFEFEF}69.13  & \cellcolor[HTML]{EFEFEF}7.39  & \cellcolor[HTML]{EFEFEF}80.10 & \cellcolor[HTML]{EFEFEF}44.03& \cellcolor[HTML]{EFEFEF}2.94  & \cellcolor[HTML]{EFEFEF}\textbf{62.96} & \cellcolor[HTML]{EFEFEF}90.59  & \cellcolor[HTML]{EFEFEF}57.49& \cellcolor[HTML]{EFEFEF}\textbf{1.11} \\
& WaNet   & 53.60 & 48.60  & 59.82  & 47.98  & 30.67 & 19.63 & 44.64 & 62.25 & 48.89 & 81.98  & 5.77  & 15.89 & 41.70  & 2.13  & \textbf{63.46} & 3.85& 58.20  & \textbf{0.00} \\
& \cellcolor[HTML]{EFEFEF}PhysicalBA & \cellcolor[HTML]{EFEFEF}71.10 & \cellcolor[HTML]{EFEFEF}100.00 & \cellcolor[HTML]{EFEFEF}\textbf{70.95} & \cellcolor[HTML]{EFEFEF}100.00 & \cellcolor[HTML]{EFEFEF}59.51 & \cellcolor[HTML]{EFEFEF}7.69  & \cellcolor[HTML]{EFEFEF}60.22 & \cellcolor[HTML]{EFEFEF}86.03 & \cellcolor[HTML]{EFEFEF}45.14 & \cellcolor[HTML]{EFEFEF}96.05  & \cellcolor[HTML]{EFEFEF}51.32 & \cellcolor[HTML]{EFEFEF}77.43 & \cellcolor[HTML]{EFEFEF}59.21& \cellcolor[HTML]{EFEFEF}0.00  & \cellcolor[HTML]{EFEFEF}61.53& \cellcolor[HTML]{EFEFEF}0.00   & \cellcolor[HTML]{EFEFEF}58.20& \cellcolor[HTML]{EFEFEF}\textbf{0.00} \\
\multirow{-6}{*}{CIFAR100}  & Average & 60.62 & 88.36  & 61.92  & 87.21  & 40.81 & 57.18 & 49.15 & 85.99 & 45.32 & 77.99  & 20.41 & 52.68 & 50.12  & 1.01  & \textbf{62.50} & 19.15  & 58.99  & \textbf{0.22} \\ \hline
& \cellcolor[HTML]{EFEFEF}BadNet  & \cellcolor[HTML]{EFEFEF}37.80 & \cellcolor[HTML]{EFEFEF}97.95  & \cellcolor[HTML]{EFEFEF}46.00& \cellcolor[HTML]{EFEFEF}98.54  & \cellcolor[HTML]{EFEFEF}26.87 & \cellcolor[HTML]{EFEFEF}97.58 & \cellcolor[HTML]{EFEFEF}31.55 & \cellcolor[HTML]{EFEFEF}90.84 & \cellcolor[HTML]{EFEFEF}5.08  & \cellcolor[HTML]{EFEFEF}84.90  & \cellcolor[HTML]{EFEFEF}0.70  & \cellcolor[HTML]{EFEFEF}0.00  & \cellcolor[HTML]{EFEFEF}32.21& \cellcolor[HTML]{EFEFEF}0.00  & \cellcolor[HTML]{EFEFEF}\textbf{46.80} & \cellcolor[HTML]{EFEFEF}0.00   & \cellcolor[HTML]{EFEFEF}35.73& \cellcolor[HTML]{EFEFEF}\textbf{0.00} \\
& Blend   & 34.60 & 99.95  & \textbf{49.20} & 100.00 & 22.29 & 96.98 & 32.11 & 99.80 & 10.37 & 18.32  & 21.54 & 99.75 & 5.33   & 0.00  & 38.65  & 0.00& 35.73  & \textbf{0.00} \\
& \cellcolor[HTML]{EFEFEF}SIG & \cellcolor[HTML]{EFEFEF}36.50 & \cellcolor[HTML]{EFEFEF}98.80  & \cellcolor[HTML]{EFEFEF}\textbf{47.26} & \cellcolor[HTML]{EFEFEF}97.79  & \cellcolor[HTML]{EFEFEF}25.52 & \cellcolor[HTML]{EFEFEF}89.43 & \cellcolor[HTML]{EFEFEF}33.87 & \cellcolor[HTML]{EFEFEF}99.39 & \cellcolor[HTML]{EFEFEF}7.60  & \cellcolor[HTML]{EFEFEF}100.00 & \cellcolor[HTML]{EFEFEF}0.75  & \cellcolor[HTML]{EFEFEF}0.00  & \cellcolor[HTML]{EFEFEF}29.89& \cellcolor[HTML]{EFEFEF}0.00  & \cellcolor[HTML]{EFEFEF}26.27& \cellcolor[HTML]{EFEFEF}78.86  & \cellcolor[HTML]{EFEFEF}38.60& \cellcolor[HTML]{EFEFEF}\textbf{0.00} \\
& WaNet   & 36.90 & 97.30  & \textbf{46.00} & 96.43  & 21.24 & 98.14 & 30.10 & 98.74 & 9.96  & 35.63  & 0.35  & 32.10 & 26.07  & 0.00  & 37.54  & 99.50  & 40.56  & \textbf{0.00} \\
& \cellcolor[HTML]{EFEFEF}PhysicalBA & \cellcolor[HTML]{EFEFEF}48.15 & \cellcolor[HTML]{EFEFEF}100.00 & \cellcolor[HTML]{EFEFEF}\textbf{52.64} & \cellcolor[HTML]{EFEFEF}100.00 & \cellcolor[HTML]{EFEFEF}33.27 & \cellcolor[HTML]{EFEFEF}33.57 & \cellcolor[HTML]{EFEFEF}42.12 & \cellcolor[HTML]{EFEFEF}93.51 & \cellcolor[HTML]{EFEFEF}12.33 & \cellcolor[HTML]{EFEFEF}20.63  & \cellcolor[HTML]{EFEFEF}1.06  & \cellcolor[HTML]{EFEFEF}98.64 & \cellcolor[HTML]{EFEFEF}29.54& \cellcolor[HTML]{EFEFEF}0.00  & \cellcolor[HTML]{EFEFEF}40.92& \cellcolor[HTML]{EFEFEF}100.00 & \cellcolor[HTML]{EFEFEF}38.70& \cellcolor[HTML]{EFEFEF}\textbf{0.00} \\
\multirow{-6}{*}{TinyImageNet} & Average & 38.79 & 98.80  & \textbf{48.22} & 98.55  & 25.84 & 83.14 & 33.95 & 96.46 & 9.07  & 51.90  & 4.88  & 46.10 & 24.61  & 0.00  & 38.04  & 55.67  & 37.86  & \textbf{0.00} \\ \hline
\end{tabular}
}
\end{table*}

\subsection{Comparative Evaluation of \texttt{MeCa} and Existing Defenses}

In this paper, we employ $8$ state-of-the-art backdoor attacks (including $5$ familiar backdoor attacks and $3$ adaptive backdoor attacks) to demonstrate the effectiveness of our \texttt{MeCa}. In this section, we take into account $5$ state-of-the-art backdoor attacks and compare the performance of \texttt{MeCa} with $7$ state-of-the-art backdoor defense techniques. Table \ref{tab:main result} shows the performance of the proposed \texttt{MeCa} method on Imagenette, CIFAR10, CIFAR100, and TinyImageNet. Obviously, our \texttt{MeCa} achieves the best results in reducing ASR against most backdoor attacks while maintaining a satisfactory ACC across all $4$ datasets. 

Concretely, we mainly have 5 observations from Table \ref{tab:main result}. The detailed analysis is shown as follows:

\textbf{(1)} Our defense method can resist to $5$ state-of-the-art backdoor attacks with an excellent performance. The ASR remains almost at $0.00\%$ on the $5$ backdoor attacks and $4$ datasets. The highest ASR is just $2.64\%$. Moreover, compared with no defense, our method has a slight drop on the main task on three datasets (CIFAR10: $87.12\%$ vs. $85.14\%$, CIFAR100:  $60.62\%$ vs. $58.99\%$, and TinyImageNet: $38.79\%$ vs. $37.86\%$). However, we find that most defense methods fail to maintain the trade-off between ACC and ASR on ImageNette. This may be attributed to the small scale of this dataset since it is sampled from a larger and more complex dataset, ImageNet, which means that the convergence of the models trained on ImageNette may be largely affected even with little data missing. With 5\% data poisoned, the original performance on 
 this dataset will also be affected. 
 
 \textbf{(2)} Apart from SCAnFT, the other $6$ state-of-the-art backdoor defense techniques perform poorly. The highest average ASR is $98.55\%$ with the FineTuning on TinyImageNet. Only for the ABL on CIFAR100, the average ASR is $19.15\%$, which is less than $20.00\%$. The average ASR of all the other defense techniques is more than $20.00\%$. For SCAnFT, it has a fine defense performance on four datasets. The average ASR is $11.41\%$, $13.65\%$, $1.01\%$, and $0.00\%$, respectively, when the SCAnFT performs on the four datasets. However, SCAnFT has an obvious influence on the main task (from $94.14\%$ to $55.13\%$, from $87.12\%$ to $70.22\%$, from $60.62\%$ to $50.12\%$, from $38.79\%$ to $24.61\%$). This indicates that SCAnFT has significant limitations when used in practical scenarios. 
 
 \textbf{(3)} The existing defense methods perform unsteadily for the same backdoor attack on the different datasets. For example, for DBR under the SIG, the ASR on four datasets is $0.28\%$, $93.52\%$, $80.10\%$, and $0.00\%$, respectively. For ABL under the Blend, the ASR on four datasets is $79.22\%$, $5.71\%$, $1.21\%$, and $0.00\%$, respectively. Since the complexity, scale, and class numbers of the different datasets vary, the convergence speed of the attack is also different. This results in a high false positive rate for poisoned sample identification when the backdoor task has not been learned. In this case, some clean samples may be picked out. Unlearning these clean samples will significantly influence the model's performance. 
 
 \textbf{(4)} All the existing defense methods have an unstable performance against different backdoor attacks. They perform satisfactorily on some backdoor attacks and may be ineffective against other backdoor attacks. This greatly reduces the availability of these defense methods in real-world scenarios because it is impractical for the defender to know the backdoor methods in advance. For example, ABL performs well on the BadNet attack but has an unsatisfactory performance on the SIG attack. Note that we have observed the different performance of these defense methods in different papers because their hyperparameters are different. We argue that it's not practical to set different hyperparameters for different attacks because defenders do not know the potential attacks. Therefore, we use the same hyperparameter configurations for different strategies, including \texttt{MeCa}.
 
 \textbf{(5)} Compared with the $7$ state-of-the-art backdoor defense techniques, our \texttt{MeCa} can achieve the lowest ASR (the highest ASR is just $2.64\%$) while maintaining a satisfactory performance on the clean samples. While some defenses reduce ASR, they do so at the cost of rendering the model practically useless, thereby failing to maintain model availability. This can be attributed to their over-unlearning characteristics to obtain low ASR, with a suboptimal trade-off between robustness and utility. Moreover, the proposed \texttt{MeCa} has a more stable performance on different datasets and various backdoor attacks.

\begin{table*}[]\centering
\caption{The comparison results (\%) between our \texttt{MeCa} and $7$ state-of-the-art backdoor defenses methods against $3$ state-of-the-art adaptive attacks on ResNet18 with $5\%$ poisoned ratio.}
\label{Table:adaptive_attack}
\renewcommand{\arraystretch}{2}
\resizebox{2\columnwidth}{!}{
\begin{tabular}{cccccccccccccccccccc}
\hline
&& \multicolumn{2}{c}{No Defense}& \multicolumn{2}{c}{FineTuning}  & \multicolumn{2}{c}{FinePruning}& \multicolumn{2}{c}{CutMix} & \multicolumn{2}{c}{CLP}& \multicolumn{2}{c}{DBR}& \multicolumn{2}{c}{SCAnFT} & \multicolumn{2}{c}{ABL} & \multicolumn{2}{c}{\texttt{MeCa}} \\ \cline{2-20} 
\multirow{-2}{*}{Dataset}& Types & ACC& ASR& ACC & ASR& ACC& ASR& ACC& ASR& ACC& ASR& ACC& ASR& ACC& ASR& ACC & ASR& ACC & ASR  \\ \hline
& \cellcolor[HTML]{EFEFEF}TaCT& \cellcolor[HTML]{EFEFEF}94.90 & \cellcolor[HTML]{EFEFEF}24.30 & \cellcolor[HTML]{EFEFEF}\textbf{72.02} & \cellcolor[HTML]{EFEFEF}3.60  & \cellcolor[HTML]{EFEFEF}61.50 & \cellcolor[HTML]{EFEFEF}4.43  & \cellcolor[HTML]{EFEFEF}69.00 & \cellcolor[HTML]{EFEFEF}21.88 & \cellcolor[HTML]{EFEFEF}57.62 & \cellcolor[HTML]{EFEFEF}9.14  & \cellcolor[HTML]{EFEFEF}66.76 & \cellcolor[HTML]{EFEFEF}9.97  & \cellcolor[HTML]{EFEFEF}59.28 & \cellcolor[HTML]{EFEFEF}0.55  & \cellcolor[HTML]{EFEFEF}50.69& \cellcolor[HTML]{EFEFEF}27.15 & \cellcolor[HTML]{EFEFEF}67.31& \cellcolor[HTML]{EFEFEF}\textbf{0.00} \\
& AdaptivaBlend & 95.80 & 70.10 & 69.81  & 11.36 & 63.16 & 29.36 & \textbf{70.36} & 2.21  & 22.99 & 21.88 & 42.11 & 36.01 & 47.37 & 12.47 & 31.02  & 55.12 & 69.25  & \textbf{0.83} \\
& \cellcolor[HTML]{EFEFEF}AdaptivaPatch & \cellcolor[HTML]{EFEFEF}93.10 & \cellcolor[HTML]{EFEFEF}94.10 & \cellcolor[HTML]{EFEFEF}70.91& \cellcolor[HTML]{EFEFEF}0.00  & \cellcolor[HTML]{EFEFEF}52.35 & \cellcolor[HTML]{EFEFEF}49.31 & \cellcolor[HTML]{EFEFEF}65.93 & \cellcolor[HTML]{EFEFEF}43.49 & \cellcolor[HTML]{EFEFEF}51.52 & \cellcolor[HTML]{EFEFEF}0.00  & \cellcolor[HTML]{EFEFEF}61.77 & \cellcolor[HTML]{EFEFEF}45.98 & \cellcolor[HTML]{EFEFEF}58.17 & \cellcolor[HTML]{EFEFEF}0.28  & \cellcolor[HTML]{EFEFEF}38.78& \cellcolor[HTML]{EFEFEF}0.00  & \cellcolor[HTML]{EFEFEF}\textbf{73.41} & \cellcolor[HTML]{EFEFEF}\textbf{0.00} \\
\multirow{-4}{*}{Imagenette}   & Average& 94.60 & 62.83 & \textbf{70.91} & 4.99  & 59.00 & 27.70 & 68.43 & 22.53 & 44.04 & 10.34 & 56.88 & 30.65 & 54.94 & 4.43  & 40.16  & 32.92 & 69.99  & \textbf{0.28} \\ \hline
& \cellcolor[HTML]{EFEFEF}TaCT& \cellcolor[HTML]{EFEFEF}82.10 & \cellcolor[HTML]{EFEFEF}63.20 & \cellcolor[HTML]{EFEFEF}82.63& \cellcolor[HTML]{EFEFEF}65.71 & \cellcolor[HTML]{EFEFEF}75.05 & \cellcolor[HTML]{EFEFEF}65.71 & \cellcolor[HTML]{EFEFEF}78.68 & \cellcolor[HTML]{EFEFEF}41.43 & \cellcolor[HTML]{EFEFEF}37.91 & \cellcolor[HTML]{EFEFEF}3.08  & \cellcolor[HTML]{EFEFEF}77.58 & \cellcolor[HTML]{EFEFEF}63.19 & \cellcolor[HTML]{EFEFEF}72.42 & \cellcolor[HTML]{EFEFEF}62.20 & \cellcolor[HTML]{EFEFEF}\textbf{85.16} & \cellcolor[HTML]{EFEFEF}48.02 & \cellcolor[HTML]{EFEFEF}82.42& \cellcolor[HTML]{EFEFEF}\textbf{0.00} \\
& AdaptivaBlend & 84.00 & 53.30 & 82.09  & 66.48 & 80.88 & 54.40 & 74.95 & 49.12 & 49.45 & 5.49  & 45.27 & 12.09 & 78.90 & 0.22  & \textbf{88.68} & 3.41  & 83.30  & \textbf{0.00} \\
& \cellcolor[HTML]{EFEFEF}AdaptivaPatch & \cellcolor[HTML]{EFEFEF}84.80 & \cellcolor[HTML]{EFEFEF}37.30 & \cellcolor[HTML]{EFEFEF}\textbf{84.07} & \cellcolor[HTML]{EFEFEF}49.78 & \cellcolor[HTML]{EFEFEF}82.42 & \cellcolor[HTML]{EFEFEF}0.11  & \cellcolor[HTML]{EFEFEF}61.65 & \cellcolor[HTML]{EFEFEF}6.48  & \cellcolor[HTML]{EFEFEF}70.33 & \cellcolor[HTML]{EFEFEF}2.09  & \cellcolor[HTML]{EFEFEF}80.22 & \cellcolor[HTML]{EFEFEF}25.27 & \cellcolor[HTML]{EFEFEF}72.64 & \cellcolor[HTML]{EFEFEF}0.00  & \cellcolor[HTML]{EFEFEF}83.30& \cellcolor[HTML]{EFEFEF}99.34 & \cellcolor[HTML]{EFEFEF}82.86& \cellcolor[HTML]{EFEFEF}\textbf{0.00} \\
\multirow{-4}{*}{CIFAR10} & Average& 83.63 & 51.27 & 82.93  & 60.66 & 79.45 & 40.07 & 71.76 & 32.34 & 52.56 & 3.55  & 67.69 & 33.52 & 74.65 & 20.81 & \textbf{85.71} & 50.26 & 82.86  & \textbf{0.00} \\ \hline
& \cellcolor[HTML]{EFEFEF}TaCT& \cellcolor[HTML]{EFEFEF}55.40 & \cellcolor[HTML]{EFEFEF}22.80 & \cellcolor[HTML]{EFEFEF}58.70& \cellcolor[HTML]{EFEFEF}25.00 & \cellcolor[HTML]{EFEFEF}42.00 & \cellcolor[HTML]{EFEFEF}17.71 & \cellcolor[HTML]{EFEFEF}53.24 & \cellcolor[HTML]{EFEFEF}17.41 & \cellcolor[HTML]{EFEFEF}18.72 & \cellcolor[HTML]{EFEFEF}32.29 & \cellcolor[HTML]{EFEFEF}41.40 & \cellcolor[HTML]{EFEFEF}29.15 & \cellcolor[HTML]{EFEFEF}44.13 & \cellcolor[HTML]{EFEFEF}0.61  & \cellcolor[HTML]{EFEFEF}\textbf{63.26} & \cellcolor[HTML]{EFEFEF}0.91  & \cellcolor[HTML]{EFEFEF}61.84& \cellcolor[HTML]{EFEFEF}\textbf{0.00} \\
& AdaptivaBlend & 55.00 & 52.90 & 58.70  & 55.26 & 40.89 & 25.20 & 42.81 & 22.47 & 37.55 & 6.17  & 2.02  & 97.87 & 57.29 & 52.83 & 62.15  & 36.64 & \textbf{63.16} & \textbf{0.00} \\
& \cellcolor[HTML]{EFEFEF}AdaptivaPatch & \cellcolor[HTML]{EFEFEF}33.40 & \cellcolor[HTML]{EFEFEF}90.70 & \cellcolor[HTML]{EFEFEF}57.29& \cellcolor[HTML]{EFEFEF}88.46 & \cellcolor[HTML]{EFEFEF}26.52 & \cellcolor[HTML]{EFEFEF}39.78 & \cellcolor[HTML]{EFEFEF}42.61 & \cellcolor[HTML]{EFEFEF}22.57 & \cellcolor[HTML]{EFEFEF}30.16 & \cellcolor[HTML]{EFEFEF}1.11  & \cellcolor[HTML]{EFEFEF}8.00  & \cellcolor[HTML]{EFEFEF}1.52  & \cellcolor[HTML]{EFEFEF}57.19 & \cellcolor[HTML]{EFEFEF}0.00  & \cellcolor[HTML]{EFEFEF}58.00& \cellcolor[HTML]{EFEFEF}0.20  & \cellcolor[HTML]{EFEFEF}\textbf{59.82} & \cellcolor[HTML]{EFEFEF}\textbf{0.00} \\
\multirow{-4}{*}{CIFAR100}  & Average& 47.93 & 55.47 & 58.23  & 56.24 & 36.47 & 27.56 & 46.22 & 20.82 & 28.81 & 13.19 & 17.14 & 42.85 & 52.87 & 17.81 & 61.14  & 12.58 & \textbf{61.61} & \textbf{0.00} \\ \hline
& \cellcolor[HTML]{EFEFEF}TaCT& \cellcolor[HTML]{EFEFEF}36.95 & \cellcolor[HTML]{EFEFEF}13.65 & \cellcolor[HTML]{EFEFEF}\textbf{46.55} & \cellcolor[HTML]{EFEFEF}10.87 & \cellcolor[HTML]{EFEFEF}26.07 & \cellcolor[HTML]{EFEFEF}8.05  & \cellcolor[HTML]{EFEFEF}32.36 & \cellcolor[HTML]{EFEFEF}19.38 & \cellcolor[HTML]{EFEFEF}2.62  & \cellcolor[HTML]{EFEFEF}0.00  & \cellcolor[HTML]{EFEFEF}0.55  & \cellcolor[HTML]{EFEFEF}0.00  & \cellcolor[HTML]{EFEFEF}25.42 & \cellcolor[HTML]{EFEFEF}0.00  & \cellcolor[HTML]{EFEFEF}40.71& \cellcolor[HTML]{EFEFEF}20.73 & \cellcolor[HTML]{EFEFEF}38.80& \cellcolor[HTML]{EFEFEF}\textbf{0.00} \\
& AdaptivaBlend & 34.75 & 73.40 & \textbf{46.00} & 74.53 & 24.81 & 35.78 & 32.91 & 56.06 & 18.52 & 20.23 & 0.20  & 77.10 & 23.45 & 0.00  & 30.20  & 52.04 & 40.82  & \textbf{0.00} \\
& \cellcolor[HTML]{EFEFEF}AdaptivaPatch & \cellcolor[HTML]{EFEFEF}38.55 & \cellcolor[HTML]{EFEFEF}24.35 & \cellcolor[HTML]{EFEFEF}\textbf{47.10} & \cellcolor[HTML]{EFEFEF}99.90 & \cellcolor[HTML]{EFEFEF}22.40 & \cellcolor[HTML]{EFEFEF}32.56 & \cellcolor[HTML]{EFEFEF}32.56 & \cellcolor[HTML]{EFEFEF}53.85 & \cellcolor[HTML]{EFEFEF}6.95  & \cellcolor[HTML]{EFEFEF}8.81  & \cellcolor[HTML]{EFEFEF}0.45  & \cellcolor[HTML]{EFEFEF}0.55  & \cellcolor[HTML]{EFEFEF}33.67 & \cellcolor[HTML]{EFEFEF}0.00  & \cellcolor[HTML]{EFEFEF}42.78& \cellcolor[HTML]{EFEFEF}99.70 & \cellcolor[HTML]{EFEFEF}39.71& \cellcolor[HTML]{EFEFEF}\textbf{0.00} \\
\multirow{-4}{*}{TinyImageNet} & Average& 36.75 & 37.13 & \textbf{46.55} & 61.77 & 24.43 & 25.46 & 32.61 & 43.10 & 9.36  & 9.68  & 0.40  & 25.88 & 27.51 & 0.00  & 37.90  & 57.49 & 39.78  & \textbf{0.00} \\ \hline
\end{tabular}
}
\end{table*}

\subsection{The Resistance to Adaptive Attacks}
To further reveal the potential risk of the \texttt{MeCa}, we also consider the adaptive adversaries that try to design special backdoor attacks to escape our \texttt{MeCa} method. The adaptive adversaries deliberately establish dependencies between the backdoor and normal functionality, which immensely increases the difficulty of backdoor sample detection. Therefore, we employ three adaptive attacks (TaCT \cite{tang2021demon}, AdaptiveBlend \cite{qi2022revisiting} and AdaptivePatch \cite{qi2022revisiting}) to demonstrate the effectiveness of our defense method. The three adaptive attacks build the dependencies between the backdoor and normal functions by using different poisoning strategies. Table \ref{Table:adaptive_attack} shows the defense results of our method and other $7$ state-of-the-art backdoor defense techniques against the $3$ adaptive attacks on $4$ different datasets. From Table \ref{Table:adaptive_attack}, we have the following three observations. First, our \texttt{MeCa} has the lowest ASR (nearly $0.00\%$) while maintaining a satisfactory ACC of the main task. Then, some existing defense methods (CLP and SCAnFT) have a good defense performance against the three adaptive attacks, but they have an obvious decrease in ACC. Specifically, for CLP on the CIFAR10, the ASR is $3.55\%$, and the ACC drops from $83.63\%$ to $52.56\%$. Finally, our method has a better generalization on different datasets. For different datasets, our method always maintains a fine performance, while the performance of the other backdoor defense techniques has an obvious fluctuation. In summary, compared with the $7$ state-of-the-art backdoor defense techniques, our method has a better performance and greater generalization against adaptive attacks. 

\subsection{Ablation Studies}
Different from the existing backdoor defense methods, we also propose to improve the performance of our \texttt{MeCa} further by relabeling and relearning the backdoor samples. Hence, in this section, we first show the \texttt{MeCa}'s performance after employing relabeling and relearning mechanisms. Then, to demonstrate our \texttt{MeCa}'s generalization, we also experimentally explore the performance of our \texttt{MeCa} on different partition rates, poison ratios, and models. Finally, we also explore the impact of our \texttt{MeCa} on the clean and poisoned dataset.
\subsubsection{Performance with Relabeling and Relearning}
In our \texttt{MeCa}, after identifying the backdoor samples and clean samples, we relabel the backdoor samples and merge them with clean samples to obtain a clean and complete dataset, which is used to fine-tune the clean model then. Accordingly, we conduct experiments on the ResNet18 and CIFAR10 datasets to show the effect of the relabeling and relearning mechanism. Table \ref{Table:relabel} shows the ACC and ASR of our \texttt{MeCa} after using relabeling and relearning. Compared with Table \ref{tab:main result} and \ref{Table:adaptive_attack}, we can know that the ACC further improves, and the ASR also decreases a little. Specifically, after employing relabeling and relearning, the average ACC and ASR changes from $84.29\%$ to $86.65\%$ and from $0.33\%$ to $0.03\%$ against the $8$ backdoor attacks, respectively. We infer that the model learns better due to the increase of clean samples. More interestingly, we find that employing relabeling and relearning has a positive effect on clean-label backdoor attacks (e.g., SIG). The ASR drops from $2.64\%$ to $0.22\%$ after equipping the relabeling and relearning mechanism. We deduce that the efficacy of backdoor task learning is closely linked to the initial weights of the model. Therefore, a pretrained model can hardly be backdoored by poisoned samples, which also indicates that model service providers use pretrained models as the base models and fine-tune them with untrusted data to avoid poisoning attacks.

\begin{table}[]\centering
\caption{The performance (\%) of ResNet18 with relabeling and complete CIFAR10 dataset.}
\label{Table:relabel}
\tabcolsep=0.42cm
\renewcommand{\arraystretch}{1.5}
\begin{tabular}{ccccc}
\hline
Stratagies→   & \multicolumn{2}{c}{\begin{tabular}[c]{@{}c@{}}w/o Relabeling\\ and Relearning\end{tabular}} & \multicolumn{2}{c}{\begin{tabular}[c]{@{}c@{}}w/ Relabeling\\ and Relearning\end{tabular}} \\ \cline{2-5} 
Attack↓& ACC & ASR  & ACC& ASR  \\ \hline
\rowcolor[HTML]{EFEFEF} 
BadNet & 85.72  & 0.00 & 86.81 & 0.00 \\
Blend  & 85.71  & 0.00 & 87.03 & 0.00 \\
\rowcolor[HTML]{EFEFEF} 
SIG & 87.69  & 2.64 & 86.59 & 0.22 \\
WaNet  & 83.63  & 0.00 & 86.92 & 0.00 \\
\rowcolor[HTML]{EFEFEF} 
PhysicalBA & 82.97  & 0.00 & 83.52 & 0.00 \\
TaCT& 82.42  & 0.00 & 86.92 & 0.00 \\
\rowcolor[HTML]{EFEFEF} 
AdaptivaBlend & 83.30  & 0.00 & 89.12 & 0.00 \\
AdaptivaPatch & 82.86  & 0.00 & 86.26 & 0.00 \\ \hline
Average& 84.29  & 0.33 & 86.65 & 0.03 \\ \hline
\end{tabular}
\end{table}

\subsubsection{Effectiveness with Different Partition Rates}
\begin{figure}
 \centering
 \includegraphics[width=1\linewidth]{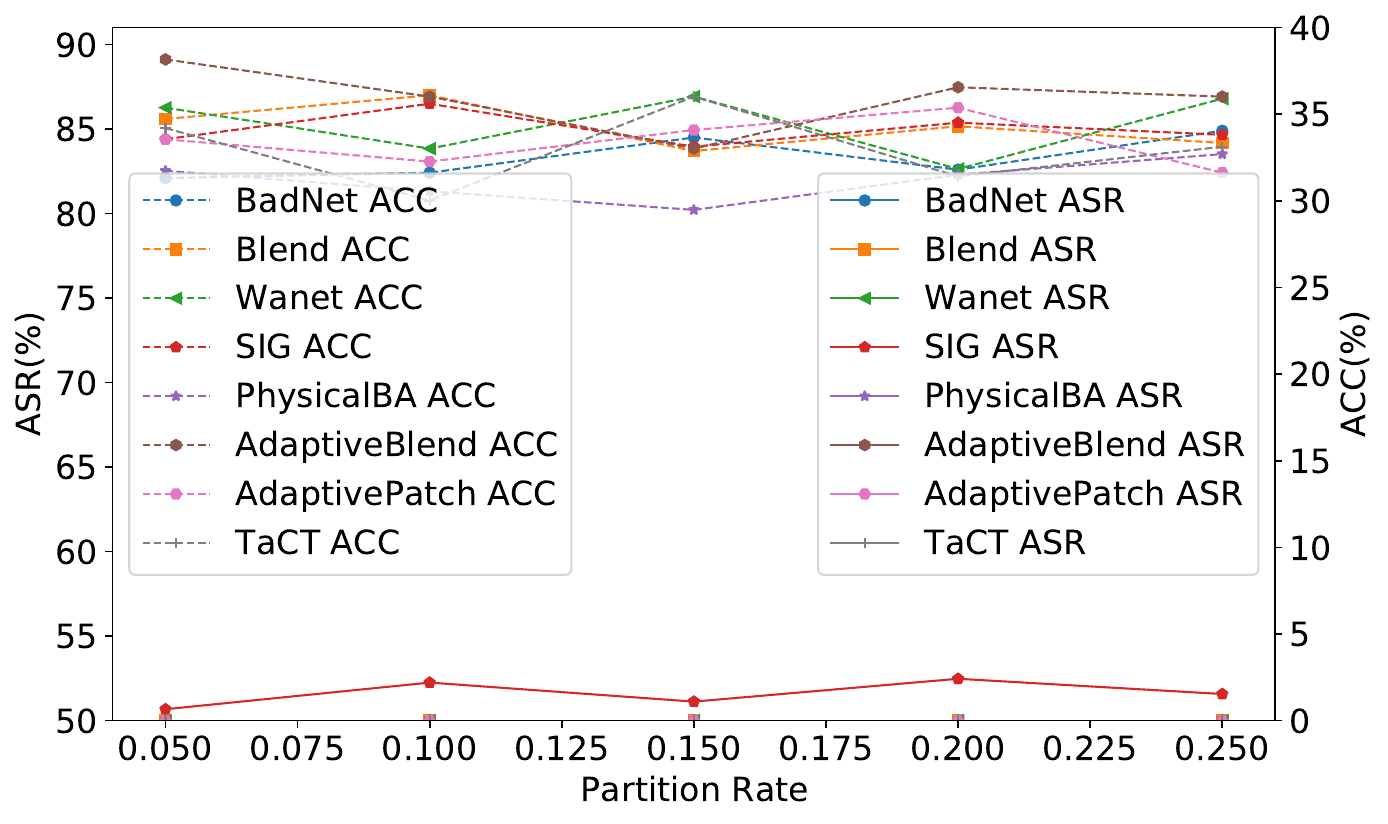}
 \caption{The ASR and ACC (\%) of \texttt{MeCa} with different partition rates.}
 \label{fig:partition_rate}
\end{figure}

In our \texttt{MeCa} method, we have to select a certain percentage of samples based on the $\mathcal{L}_{kl}$ value to train a backdoor enhancement model. This means that the partition rate may be important for \texttt{MeCa}'s performance. Therefore, to explore the effect of partition rate on performance, we conduct experiments on CIFAR10 and $8$ backdoor attacks when the partition rate is $0.05$, $0.10$, $0.15$, $0.20$, and $0.25$, respectively. The experimental results are shown in Fig.\ref{fig:partition_rate}, which shows that the partition rates have little influence on the \texttt{MeCa}'s performance. Our method on all the backdoor attacks apart from SIG has a stable performance. The ASR is always nearly $0.00\%$ except for SIG, in which the ASR causes a little fluctuation with a very small margin around $0.00\%$. We infer that it's a normal fluctuation because of the occasionality of each experiment. For ACC, there is also a little fluctuation around $85.00\%$ for all the backdoor attacks. We infer that it is also a normal fluctuation due to the occasionality of each experiment.

\begin{table}[t]\centering
\caption{Impact of poison ratio on defense performance (\%).}
\label{Table:poison rate}
\renewcommand{\arraystretch}{1.5}
\begin{threeparttable}
\begin{tabular}{ccccccc}
\hline
Poison Ratio→ & \multicolumn{2}{c}{10\%} & \multicolumn{2}{c}{15\%} & \multicolumn{2}{c}{20\%} \\ \cline{2-7} 
Method↓& ACC  & ASR& ACC  & ASR & ACC  & ASR\\ \hline
\rowcolor[HTML]{EFEFEF} 
BadNet & 85.82& 0.00 & 85.38& 0.00& 82.20& 0.00 \\
Blend  & 82.00& 0.00 & 83.30& 0.00& 82.20& 0.00 \\
\rowcolor[HTML]{EFEFEF} 
WaNet  & 83.74& 0.00 & 85.05& 0.00& 82.42& 0.00 \\
PhysicalBA & 82.42& 0.00 & 81.76& 0.00& 81.10& 0.00 \\
\rowcolor[HTML]{EFEFEF} 
TaCT& 78.35& 0.00 & 78.90& 0.00& 81.10& 0.00 \\
AdaptiveBlend & 87.90& 0.00 & 87.69& 0.00& 78.35& 0.00 \\
\rowcolor[HTML]{EFEFEF} 
AdaptivePatch & 87.58& 0.00 & 87.25& 0.00& 86.26& 0.00 \\ \hline
Average& 83.50& 0.00 & 83.87& 0.00& 81.98& 0.00 \\ \hline
\end{tabular}
 \begin{tablenotes}
  \item $\bullet$ Without attack: ACC-86.04\%, ASR-0.66\%
   \end{tablenotes}
\end{threeparttable}

\end{table}

\subsubsection{Effectiveness with Different Poison Ratios}
In real-world application scenarios, it is impractical for the defender to access the poison ratios of the training data. Therefore, we also demonstrate that our \texttt{MeCa} is suitable for multiple poison ratios. Here, we experiment it on CIFAR10 against $7$ backdoor attacks, including BadNet, Blend, WaNet, PhysicalBA, TaCT, AdaptivaBlend, and AdaptivaPatch with poison ratios up to $0.1$, $0.15$, $0.2$. Note that SIG is not considered here because it is a clean-label attack that can just poison one-class samples. Moreover, the number of one-class samples is a maximum of $10.00\%$ in CIFAR10. That is, the poison ratios of SIG on CIFAR10 do not exceed $10.00\%$. Table \ref{Table:poison rate} shows the experimental results. From Table \ref{Table:poison rate}, we can know that with the variation of poison ratios, the ASR of our method is always $0.00\%$ against the $7$ backdoor attacks. In particular, we can find that the ASR is $0.66\%$ when the poison ratio is zero. We suppose that the reason is the misclassification of the model on the clean samples. In addition, the experimental results show that our method also maintains a fine ACC against different backdoor attacks even without employing relabeling and relearning methods.

\subsubsection{Effectiveness with Different Models}
To demonstrate the generalizability of our method, we conduct experiments on ResNet34 and MobileNetV2 with the poison ratio of $5\%$. Table \ref{Table:models} shows the experimental results on CIFAR10. From Table \ref{Table:models}, we can find that our \texttt{MeCa} works similarly well on ResNet34 and MobileNetV2. The ASR is always $0.00\%$ in the two model architectures. Moreover, the ACC also maintains a satisfactory performance (from $88.60\%$ to $88.70\%$ in ResNet34, from $84.33\%$ to $81.89\%$ in MobileNetV2). The experimental results indicate that our \texttt{MeCa} has good generalizability between different models.

\begin{table}[t]\centering
\caption{The performance (\%) of ResNet34 and MobileNetV2 on CIFAR10.}
\label{Table:models}
\tabcolsep=0.42cm
\renewcommand{\arraystretch}{1.5}
\begin{tabular}{ccccc}
\hline
Model→ & \multicolumn{2}{c}{ResNet34} & \multicolumn{2}{c}{MobileNetV2} \\ \cline{2-5} 
Attack↓& ACC & ASR& ACC   & ASR \\ \hline
\rowcolor[HTML]{EFEFEF} 
No Attack  & 88.60  & 0.00  & 84.33 & 0.00\\
BadNet & 88.35  & 0.00  & 80.30 & 0.00\\
\rowcolor[HTML]{EFEFEF} 
Blend  & 88.35  & 0.00  & 81.76 & 0.00\\
SIG & 89.56  & 0.00  & 82.31 & 0.00\\
\rowcolor[HTML]{EFEFEF} 
WaNet  & 88.35  & 0.00  & 81.20 & 0.00\\
PhysicalBA & 87.69  & 0.00  & 80.21 & 0.00\\
\rowcolor[HTML]{EFEFEF} 
TaCT& 89.45  & 0.00  & 82.63 & 0.00\\
AdaptiveBlend & 89.01  & 0.00  & 80.22 & 0.00\\
\rowcolor[HTML]{EFEFEF} 
AdaptivePatch & 88.90  & 0.00  & 84.07 & 0.00\\ \hline
Average& 88.70  & 0.00  & 81.89 & 0.00\\ \hline
\end{tabular}
\end{table}

\begin{table}[t]
\centering
\caption{Impact of \texttt{MeCa} on the poisoned dataset.}
\label{Table:defense_vs_clean}
\tabcolsep=0.42cm
\renewcommand{\arraystretch}{1.5}
\begin{threeparttable}
\begin{tabular}{ccccc}
\hline
& \multicolumn{2}{c}{Poison w/o defense} & \multicolumn{2}{c}{Poison w/ defense} \\ \cline{2-5} 
\multirow{-2}{*}{Attack Method} & ACC  & ASR & ACC & ASR \\ \hline
\rowcolor[HTML]{EFEFEF} 
BadNet  & 92.00   & 95.20 & 85.72 & 0.00   \\
Blended & 84.50   & 99.90 & 85.71 & 0.00   \\
\rowcolor[HTML]{EFEFEF} 
SIG  & 84.70   & 95.40 & 87.69 & 2.64   \\
WaNet& 82.80   & 51.30 & 83.63 & 0.00   \\
\rowcolor[HTML]{EFEFEF} 
PhysicalBA  & 91.60   & 100.00   & 82.97 & 0.00   \\
TaCT & 82.10   & 63.20 & 82.42 & 0.00   \\
\rowcolor[HTML]{EFEFEF} 
AdaptivaBlend  & 84.00   & 53.30 & 83.30 & 0.00   \\
AdaptivaPatch  & 84.80   & 37.30 & 82.86 & 0.00   \\ \hline
Average & 85.81   & 74.45 & 84.29 & 0.33   \\ \hline
\end{tabular}
\begin{tablenotes}
 \item $\bullet$ Clean dataset without defense: ACC-87.86\%, ASR-0.00\%
 \item $\bullet$ Clean dataset with defense: ACC-86.04\%, ASR-0.66\%
\end{tablenotes}
\end{threeparttable}
\end{table}

\subsubsection{Impact on clean and poisoned dataset}
In real-world applications, it is impractical for the defender to know whether a given training dataset contains some poisoned samples. Therefore, to demonstrate the impact of our \texttt{MeCa} on the clean and poisoned dataset, we conduct experiments on CIFAR10 with the poison ratio of $0$ and $5\%$. The experimental results (Table \ref{Table:defense_vs_clean}) illustrate that for training on the poisoned dataset, our \texttt{MeCa} can significantly reduce the average ASR (from $74.45\%$ to $0.33\%$) with a slight loss of ACC (from $85.81\%$ to $84.29\%$). On the other hand, when trained on a clean dataset, our \texttt{MeCa} has a minor impact on ACC (from $87.86\%$ to $86.04\%$). Note that the ASR is $0.66\%$ when training on the clean dataset with our \texttt{MeCa}. We infer that this can be attributed to the misclassification of the model on the clean samples. The experimental results further demonstrate that \texttt{MeCa} is a practical solution for dealing with unknown datasets.

\section{Conclusion}
In this paper, we first explore the relationship between the backdooor and perturbation by our theoretical analysis and experimental verification. Based on the results of the investigation, we propose a novel backdoor defense method (\texttt{MeCa}) to identify poisoned samples and train a clean model on a poisoned dataset. The proposed \texttt{MeCa} partitions the poisoned samples and clean samples according to their robustness to adversarial perturbation. There is no requirement for any auxiliary clean dataset or knowledge about the poisoned dataset (e.g., poison ratios) in the \texttt{MeCa}. Extensive experimental results show the superior performance of \texttt{MeCa} in defending against $8$ state-of-the-art backdoor attacks. Compared with the $7$ advanced backdoor defense methods, our \texttt{MeCa} has a lower ASR while maintaining a satisfactory ACC on the main task. In addition, the experimental results also demonstrate that the proposed \texttt{MeCa} has a fine generalization ability in different poison ratios and various model architectures. The relevant experimental results indicate our method has prominent potential and vital practicality for real-world application scenarios as well.

\bibliographystyle{IEEEtran}
\bibliography{main}

\begin{IEEEbiography}
[{\includegraphics[width=1in,height=1.25in,clip,keepaspectratio]{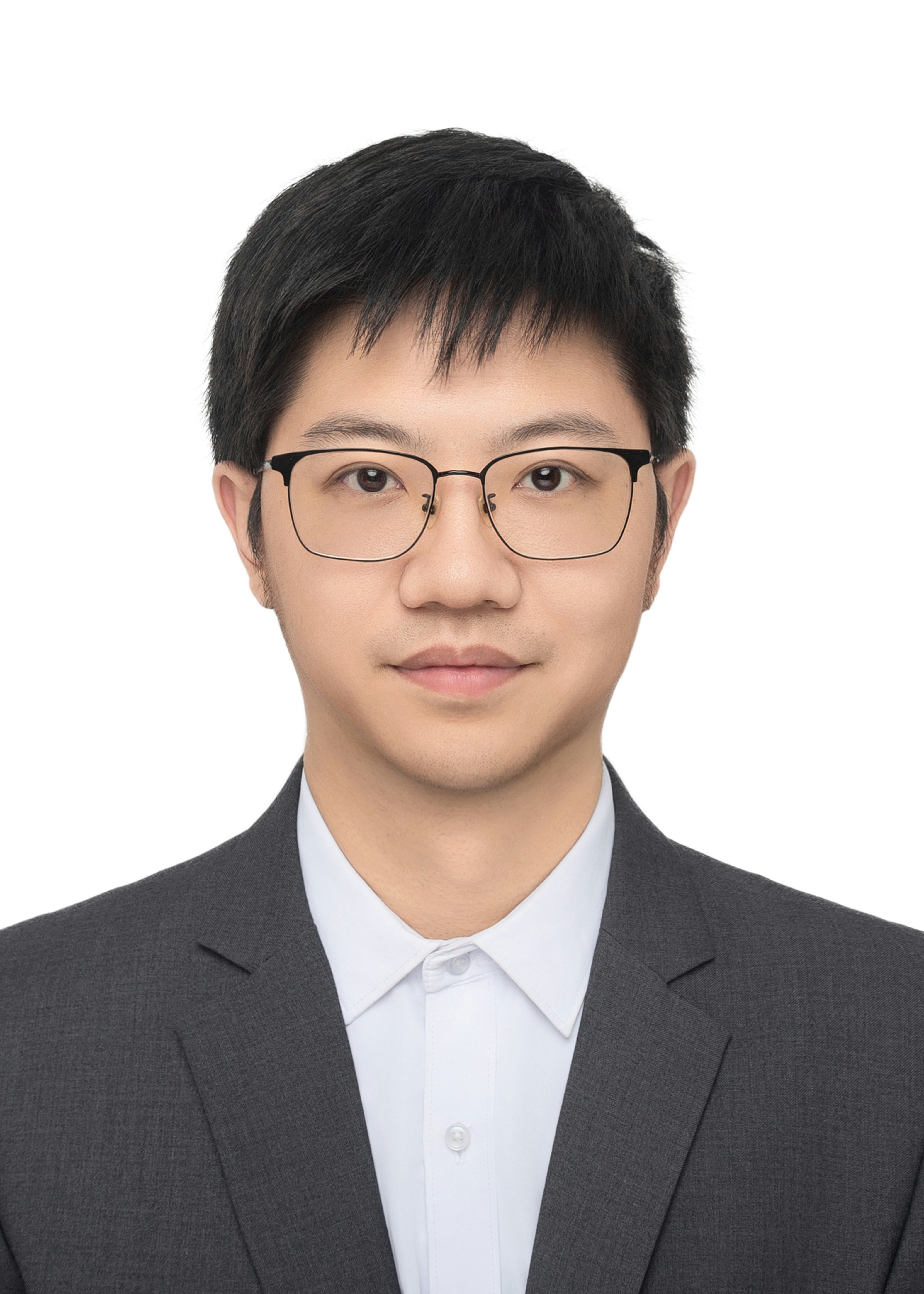}}] 
{Yuwen Pu} received his Ph.D. in the School of Big Data \& Software Engineering from Chongqing University in 2021. He is an associate professor at the School of Big Data \& Software Engineering at Chongqing University. His research interests include big data security and privacy-preserving, AI security.
\end{IEEEbiography}

\vspace{-0.5\baselineskip}
\begin{IEEEbiography}
[{\includegraphics[width=1in,height=1.25in,clip,keepaspectratio]{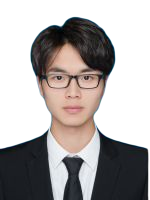}}]
{Jiahao Chen} received his B.S. at the Faculty of Electrical Engineering and Computer Science in Ningbo University in 2023. He is pursuing his Ph.D. degree in the College of Computer Science and Technology at Zhejiang University. His research interests include federated learning, privacy-preserving, and AI security.
\end{IEEEbiography}

\vspace{-0.5\baselineskip}
\begin{IEEEbiography}
[{\includegraphics[width=1in,height=1.25in,clip,keepaspectratio]{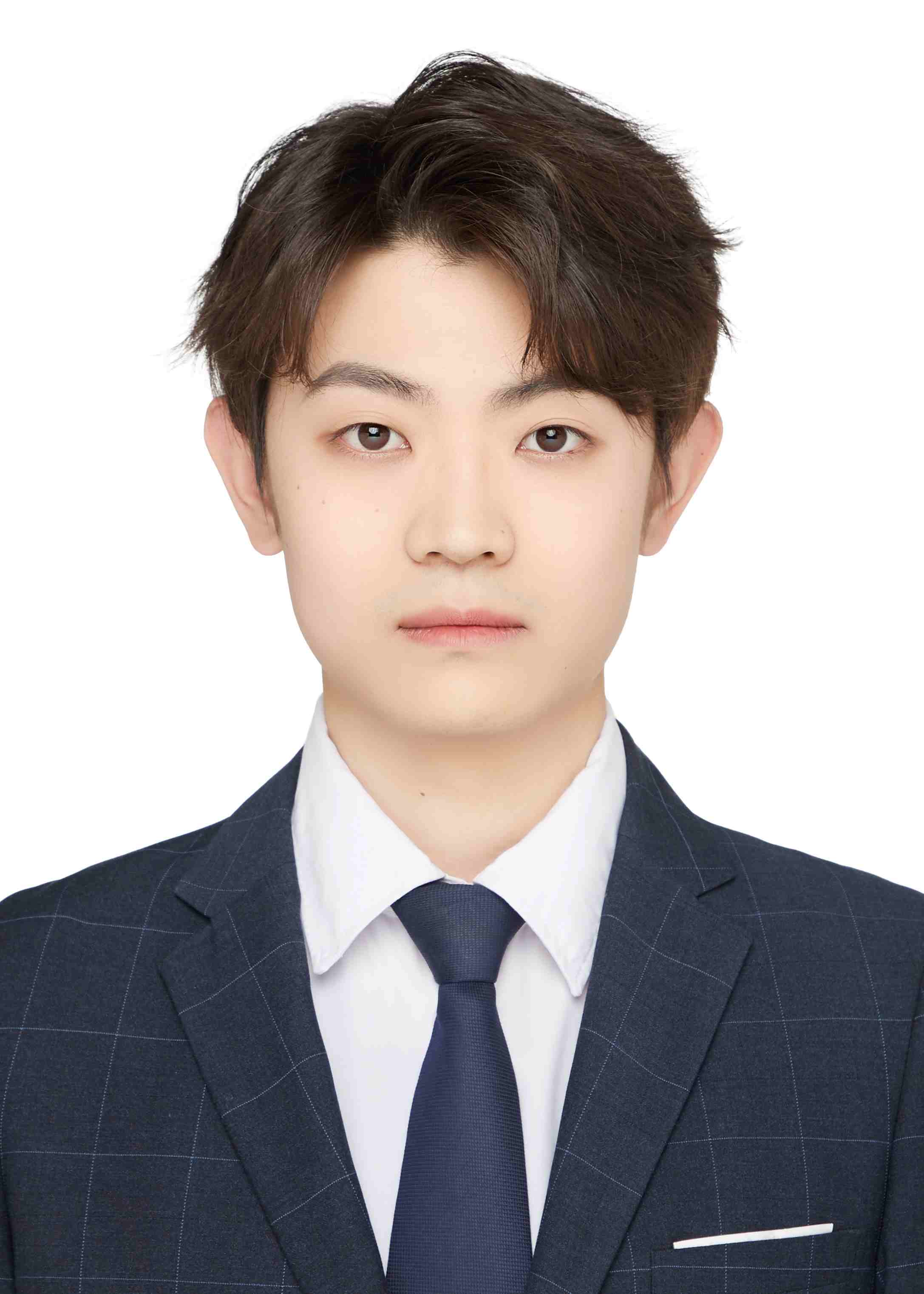}}]
{Chunyi Zhou} received the PhD degree with the School of Computer Science and Engineering from Nanjing University of Science and Technology in 2024  (supervised by Prof. Anmin Fu). He is currently a PostDoctor collaborating with Prof. Shouling Ji at the College of Computer Science and Technology, Zhejiang University. His research interests include AI privacy and security, federated learning, and machine unlearning.
\end{IEEEbiography}

\vspace{-0.5\baselineskip}
\begin{IEEEbiography}
[{\includegraphics[width=1in,height=1.25in,clip,keepaspectratio]{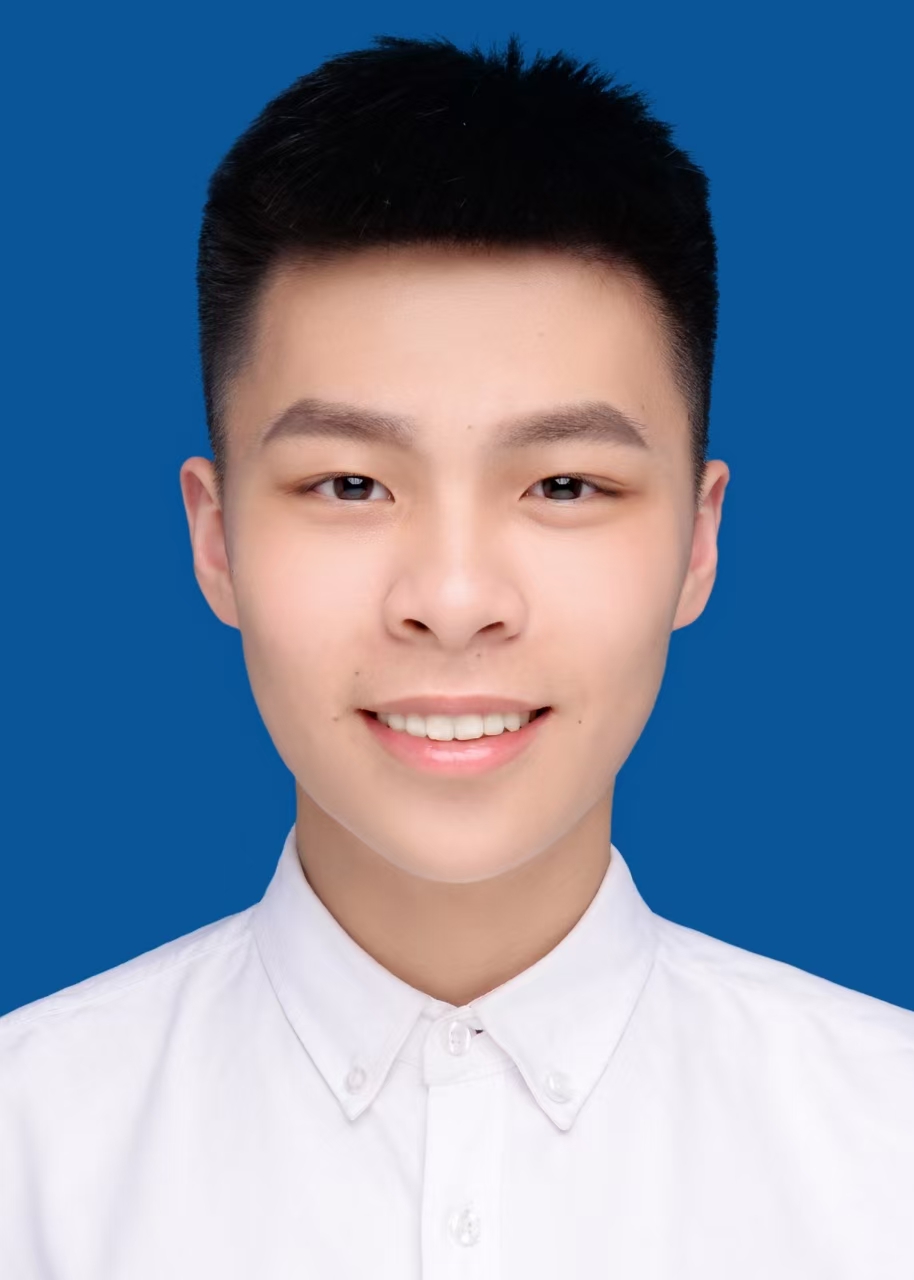}}]
{Zhou Feng} is a graduate student at the College of Computer Science and Technology, Zhejiang University, and a member of the NESA Lab. His research interests lie in backdoor attacks and audio privacy protection, with a focus on adversarial machine learning and secure AI systems.
\end{IEEEbiography}

\vspace{-0.5\baselineskip}
\begin{IEEEbiography}
[{\includegraphics[width=1in,height=1.25in,clip,keepaspectratio]{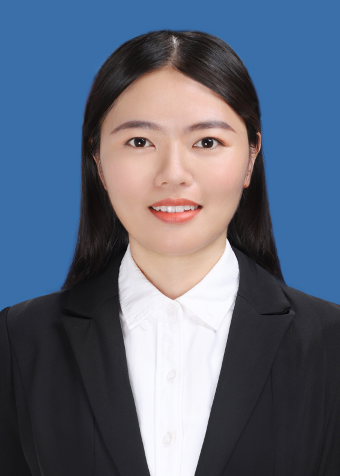}}]
{Qingming Li} received the B.S. degree from Xidian University, Xi'an, China, in 2017, and the Ph. D degree from Tsinghua University, Beijing, China, in 2022. She was a Research Assistant with the Research Institute of Artificial Intelligence, Zhejiang Lab, Hangzhou, China from July 2022 to Mar. 2024. Since Apr. 2024, she has been a Postdoc with the College of Computer Science and Technology, Zhejiang University, Hangzhou, China. Her research interests include deep learning security, large language models, federated learning, and data valuation. 
\end{IEEEbiography}

\vspace{-0.5\baselineskip}
\begin{IEEEbiography}
[{\includegraphics[width=1in,height=1.25in,clip,keepaspectratio]{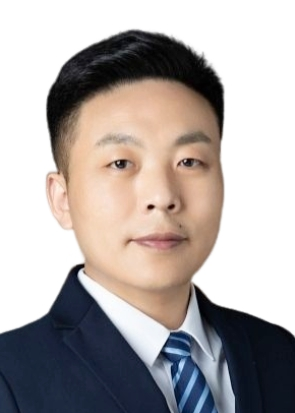}}]
{Chunqiang Hu} received the BS degree in computer science and technology from Southwest University, Chongqing, China, in 2006, the MS and PhD degrees in computer science and technology from Chongqing University, Chongqing, China, in 2009 and 2013, respectively, and the second PhD degree in computer science from the George Washington University, Washington, DC, USA, in 2016. He was a visiting scholar with the George Washington University from 2011 to 2012. He is currently a professor with the School of Big Data \& Software Engineering, Chongqing University. His research interests include Blockchain, privacy computing, Data-Driven Security, applied cryptography, and algorithm design and analysis. He was honored with the Hundred-Talent Program by Chongqing University. He is a senior member of the CCF (China Computer Federation).
\end{IEEEbiography}

\vspace{-0.5\baselineskip}
\begin{IEEEbiography}
[{\includegraphics[width=1in,height=1.25in,clip,keepaspectratio]{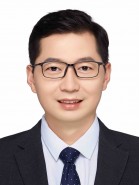}}]
{Shouling Ji} is a Professor in the College of Computer Science and Technology at Zhejiang University and a Research Faculty in the School of Electrical and Computer Engineering at Georgia Institute of Technology. He received a Ph.D. in Electrical and Computer Engineering from Georgia Institute of Technology, a Ph.D. in Computer Science from Georgia State University. His current research interests include AI Security, Data-driven Security, Privacy and Data Analytics. He is a member of IEEE and ACM and was the Membership Chair of the IEEE Student Branch at Georgia State (2012- 2013).
\end{IEEEbiography}

\end{document}